\documentclass[a4paper,twocolumn,11pt,accepted=2017-05-09]{quantumarticle}
\makeatletter
\newsavebox\@quantumacceptedbox
\makeatother
\pdfoutput=1
\usepackage[utf8]{inputenc}
\usepackage[english]{babel}
\usepackage[T1]{fontenc}
\usepackage{amsmath}
\usepackage{amssymb}
\usepackage{hyperref}

\usepackage{tikz}
\usepackage{lipsum}

\begin{document}

\title{Topological Winding Readout of an Emergent Page-Wootters Clock}

\author{Hesam Zaravashan}
\author{Gabriele Gradoni}
\author{Mohsen Khalily}
\affiliation{Institute for Communication Systems (ICS), University of Surrey, Guildford, GU2 7XH, United Kingdom}
\maketitle

\begin{abstract}
The Page--Wootters construction gives time a Hilbert space, representing it as
an internal degree of freedom whose readings the rest of the system evolves with
respect to.  We ask whether such a clock degree of freedom can carry a
topological invariant, and we propose a photonic architecture in which it does
and in which the invariant can be measured.  Signal--idler pairs from
spontaneous four-wave mixing in a coupled microring array supply the
clock--system partition, with the idler occupying a Rice--Mele band whose
staggered coupling and detuning are cycled adiabatically around a gap-closing
point.  One cycle acts on that band as a rigid translation by \(C\) unit cells,
times a geometric phase periodic in the crystal momentum, times a dynamical
phase, an identity exact in the adiabatic limit and independent of the state on
which it acts.  The readout targets the winding of that geometric phase across
the Brillouin zone.  Reversing the traversal of the pump loop cancels the
dynamical phase and isolates twice the geometric phase in a two-photon
coincidence fringe, while energy anticorrelation lets a filter on the signal
scan the clock momentum without touching the clock and supplies the phase
reference relationally, in place of an external optical reference.  The estimate
is protected on two independent levels, by the gap against smooth deformation
and by the digital character of the unwrapping against noise below a threshold.
A parameter budget anchored to thin-film lithium niobate places the requirements
within reach at a conservative operating point.
\end{abstract}

%
%
%
%

\section{Introduction}
\label{sec:introduction}

Time in ordinary quantum mechanics is an external parameter, a label for
evolution rather than a property of the system itself.  What the Page--Wootters
construction supplies is a Hilbert space for
it~\cite{page1983evolution,hohn2021trinity}: a global state on \(\mathcal
H_C\otimes\mathcal H_S\) satisfies a Hamiltonian constraint and encodes
correlations between a clock \(C\) and a system \(S\), so that conditioning on a
clock reading recovers the ordinary Schr\"odinger evolution of the system.  This
has been illustrated with entangled photons~\cite{moreva2014time} and extended
to interacting clock--system pairs~\cite{smith2019quantizing}.  Once time is
carried by a Hilbert space, a question opens that could not previously be
posed at all: whether a clock degree of freedom defined this way carries a
topological invariant, and whether that invariant can be measured.

Neither half of the question is answered by the work that surrounds it.
Symmetry-protected clock spectroscopy uses a winding to protect the coherence
of atoms whose evolution is still referred to an external
time~\cite{xu2025symmetry}, while Berry connections enter relational and
cosmological clock models as gauge terms in the conditional dynamics, neither
quantised nor measured~\cite{nambu2022qubit}.  Interferometric measurement of a
Zak phase is by now standard~\cite{atala2013direct}, but the phase there belongs
to an ordinary Bloch band and is read against an external reference.  On the
topological side the ingredients are equally well established: integer
invariants survive smooth deformation as long as a gap stays
open~\cite{xiao2010berry,hasan2010colloquium}, the Thouless pump is the
canonical dynamical example~\cite{thouless1983quantization,citro2023thouless},
and the winding of a Wilson loop across the Brillouin zone is the spectral flow
of hybrid Wannier charge centres, a standard route to the Chern
number~\cite{soluyanov2011computing,gresch2017z2pack}.  What has not been done
is to attach that structure to a relational clock and to read it out without an
external optical phase reference, which is what this paper proposes.

The platform is a one-dimensional array of coupled microring
resonators~\cite{dutt2020single,yuan2018synthetic,mittal2018topological,zhang2017monolithic},
in which a degenerate pump generates signal--idler pairs by spontaneous
four-wave mixing.  Taking the idler as the clock and the signal as the system,
the clock Hilbert space is a single band seen by the idler, described by a
Rice--Mele model~\cite{rice1982elementary} whose staggered coupling and
staggered detuning are cycled adiabatically around a gap-closing point.  A band
alone is not a clock; what makes it one is the covariant family of time states
built on it~\cite{holevo1982probabilistic,busch1994time,loveridge2019relative},
for which evolution by any duration translates the reading rigidly.  The reading
is the label \(t\) of that family, not the crystal momentum \(k_x\) and not the
position of the photon along the chain, and because a band Hamiltonian is
bounded the family is a covariant positive-operator-valued measure rather than a
self-adjoint time operator, which no such Hamiltonian admits.  Distinct from all
of this is the slow control \(\tau\) that drives the pump cycle, a classical
parameter tracing a closed loop in the space of clock Hamiltonians: one is read
off the clock photon, the other is set by the modulation electronics.  Since the
topological content of a closed loop does not depend on how the loop is
parametrised, no external time scale enters the invariant.

Because the Berry phase \(\gamma(k_x)\) accumulated over the closed cycle winds
by exactly \(2\pi\) per Chern unit across the zone, and a phase linear in
momentum is a translation, one cycle acts on the clock band as a rigid
translation by \(C\) unit cells, times a geometric phase periodic in \(k_x\),
times a dynamical phase.  This identity is exact in the adiabatic limit and
holds for the entire band, with no assumption about the dispersion and none
about the state on which it acts.  Its quantised content is a translation in
space, of the lattice on which the emergent time is defined, familiar from
Thouless pumping and the theory of
polarisation~\cite{kingsmith1993theory,resta1998quantum}.

That translation, however, acts in full only on a state of uniform weight across
the zone, since the quantisation comes from the zone integral of the Berry
curvature weighted uniformly, which a narrow momentum window does not
provide~\cite{thouless1983quantization,ke2020topological}, and a narrowband
heralded source does not supply such a state: the packet it prepares follows the
local Berry curvature and is displaced by a non-quantised
amount~\cite{lu2016geometrical,wimmer2017experimental}.  The winding of
\(\gamma(k_x)\) across the zone carries the same integer without that
requirement, so it is the representation the readout below uses.  Extracting it
means isolating the geometric phase from the dynamical one, and traversing the
pump loop backward does exactly that: the orientation of the parameter loop
reverses while the dynamical integral is unchanged, so the fringe difference
between forward and backward traversals in a two-photon coincidence
interferometer returns \(2\gamma(k_x)\), a separation already demonstrated in a
superconducting charge pump~\cite{mottonen2008experimental}.  The clock momentum
is then scanned through the partner rather than through the clock, since energy
anticorrelation lets a narrowband filter on the signal herald the idler at a
definite \(k_x\) while the signal amplitude supplies the phase reference of the
fringe.  Unwrapping across the zone returns the integer \(2C\), up to the
orientation convention fixed in Sec.~\ref{sec:cycle_chern}.  Here the
entanglement is operational rather than interpretational: an interferometric
phase exists only against a reference, and the partner photon supplies it
relationally, in place of an external optical standard.

Protection here comes in two layers that are logically independent of one
another, the first residing in the band structure and the second in the
estimator.  The gap pins the ideal winding
against smooth deformations of the device and of the pump loop, so the integer
moves only through a gap closing.  The estimator, for its part, is digital,
rounding the unwrapped fringe to the nearest integer, so below a threshold set
by the sampling margin the misidentification probability falls exponentially,
where an analog delay estimate degrades linearly in the same noise.  Both
statements are given below as inequalities with their conditions attached.

The paper is organised as follows.  Section~\ref{sec:photonic_paw_clock} defines
the Rice--Mele clock band, the signal--idler Page--Wootters partition, and the
covariant clock states.  Section~\ref{sec:topological_cycle} derives the Chern
number of the cycle and proves the factorisation.
Section~\ref{sec:winding_readout} constructs the forward--backward
interferometer and the nonlocal momentum scan.
Section~\ref{sec:protection_feasibility} treats gap protection, sampling, and
readout errors, and Sec.~\ref{sec:feasibility} the parameter budget.
Section~\ref{sec:conclusion} concludes, and the appendices collect the
derivations and numerical checks.


\section{Photonic Page--Wootters clock}
\label{sec:photonic_paw_clock}
 
\subsection{Rice--Mele microring chain}
\label{sec:rice_mele_clock_band}

\begin{figure*}[t]
\centering
\includegraphics[width=\textwidth]{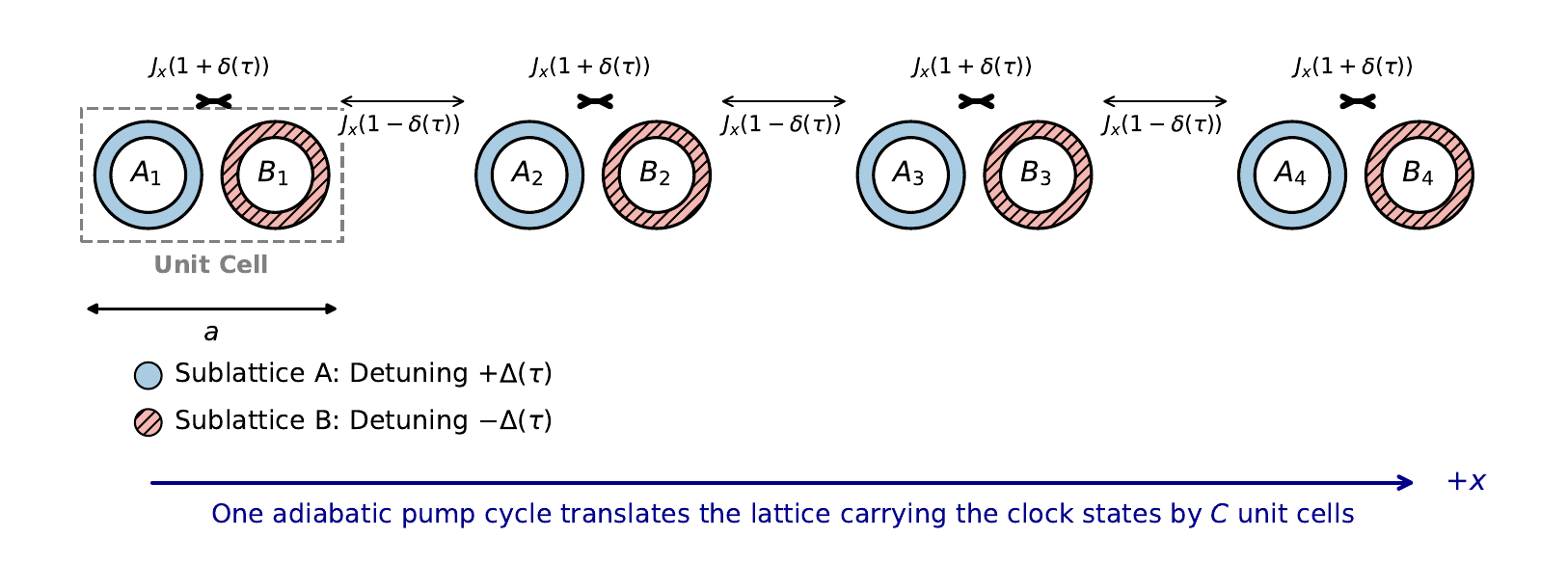}
\caption{%
Rice--Mele microring chain carrying the clock band.  Each unit cell contains two
rings, \(A\) and \(B\), of length \(a\).  Alternating couplings \(J_x(1\pm\delta)\) dimerise the chain and a staggered
detuning \(\pm\Delta\) offsets the two sublattice resonances; both are control
parameters, cycled adiabatically around the Dirac point at
\((\delta,\Delta)=(0,0)\) in Sec.~\ref{sec:topological_cycle}.
}
\label{fig:figure1}
\end{figure*}
 
The platform is a one-dimensional array of coupled microring resonators with a two-ring unit cell of length \(a\) (Fig.~\ref{fig:figure1}), a standard setting for topological band engineering in photonics~\cite{ozawa2019topological}.  The two sublattices are denoted by
\(A\) and \(B\), and the unit-cell coordinate is the spatial direction along the
array.  Alternating nearest-neighbour couplings \(J_x(1+\delta)\) and
\(J_x(1-\delta)\) dimerise the chain, while a staggered electro-optic detuning
\(\pm\Delta\) raises and lowers the resonance frequencies of the two sublattices.
In the Bloch basis \((A,B)\), the spatial Hamiltonian is the Rice--Mele
model~\cite{rice1982elementary}
\begin{equation}
\label{eq:rice_mele_hamiltonian}
\begin{aligned}
H_{\mathrm{sp}}(k_x;\delta,\Delta)
={}&
d_x(k_x,\delta)\sigma_x
+d_y(k_x,\delta)\sigma_y
\\
&+d_z(\Delta)\sigma_z
+\omega_0 I ,
\end{aligned}
\end{equation}
with
\begin{align}
\label{eq:rice_mele_components}
d_x(k_x,\delta)
&=J_x\left[(1+\delta)+(1-\delta)\cos k_x\right],
\\
d_y(k_x,\delta)
&=J_x(1-\delta)\sin k_x,
\\
d_z(\Delta)&=\Delta.
\end{align}
Here \(k_x\in(-\pi,\pi]\) is the dimensionless crystal momentum of the chain, and
\(J_x\), \(\Delta\), and \(\omega_0\) have units of angular frequency, while
\(\delta\) is dimensionless.

The lower band of Eq.~\eqref{eq:rice_mele_hamiltonian} has instantaneous
frequency
\begin{equation}
\label{eq:band_dispersion}
\omega(k_x;\delta,\Delta)=\omega_0-\sqrt{d_x^2+d_y^2+d_z^2}
\end{equation}
at fixed control parameters.  The radicand admits a closed form that will be used
repeatedly,
\begin{equation}
\label{eq:dispersion_closed_form}
d_x^2+d_y^2+d_z^2
=
\Delta^2
+
2J_x^2\left[1+\delta^2+(1-\delta^2)\cos k_x\right],
\end{equation}
from which two properties follow at once. The dispersion is even in \(k_x\), and
on the half zone \(k_x\in(0,\pi)\) it is strictly monotonic for every
\(|\delta|<1\), with the group velocity vanishing only at the two zone edges
\(k_x=0,\pi\), which are band extrema for all values of the control parameters.
 
The tight-binding description assumes that, near the resonances, each ring contributes a single longitudinal mode to the spatial
chain. Coupling to adjacent longitudinal resonances is perturbative when \(J_x\)
is small compared with the free spectral range; the leading correction scales as
\((J_x/\mathrm{FSR})^2\). For the parameter budgets discussed in
Sec.~\ref{sec:protection_feasibility}, this correction renormalises the
dispersion while leaving the gap open, and the winding is unchanged.

 
\subsection{Signal--idler Page--Wootters partition}
\label{sec:sfwm_paw_partition}
 
The clock--system partition is supplied by a signal--idler photon pair generated
through spontaneous four-wave mixing. A degenerate pump at frequency
\(\omega_p\) drives the third-order nonlinearity and creates one signal photon
and one idler photon subject to energy conservation,
\begin{equation}
\label{eq:sfwm_energy_conservation}
\omega_s+\omega_i=2\omega_p.
\end{equation}
Throughout, the idler is the clock \(C\) and the signal is the system \(S\); the two names are used interchangeably. The single-pair Hilbert space therefore factorises as
\begin{equation}
\label{eq:hilbert_factorisation}
\mathcal H=\mathcal H_C\otimes\mathcal H_S.
\end{equation}
The signal and idler occupy two longitudinal modes of the same chain, separated
by several free spectral ranges.  Each therefore has a band of the form of
Eq.~\eqref{eq:band_dispersion}, with its own carrier offset and its own effective
parameters, since the couplers are frequency dependent.  The clock Hamiltonian
\(H_C\) acts on the idler band, with dispersion \(\omega_C(k_x)\), and the system
Hamiltonian \(H_S\) on the signal band, with dispersion \(\omega_S(k_x)\).
 
The pair produced by spontaneous four-wave mixing is not a product state. In the
band basis it has the form
\begin{equation}
\label{eq:sfwm_state}
\begin{aligned}
|\Psi\rangle\rangle
={}&
\sum_{k_x^{(i)},k_x^{(s)}}
\Phi\!\left(k_x^{(i)},k_x^{(s)}\right)
\\
&\times
|u_C(k_x^{(i)})\rangle
\otimes
|u_S(k_x^{(s)})\rangle ,
\end{aligned}
\end{equation}
where \(k_x^{(i)}\) and \(k_x^{(s)}\) are the Bloch momenta of the idler and the signal. Phase matching and energy conservation make
\(\Phi\) sharply peaked near the anti-correlation manifold
\begin{equation}
\label{eq:energy_anticorrelation}
\omega_C\!\left(k_x^{(i)}\right)+\omega_S\!\left(k_x^{(s)}\right)
\simeq 2\omega_p.
\end{equation}
This anti-correlation is the resource that allows conditioning on a clock
reading to return a definite conditional state of the signal
photon~\cite{page1983evolution,moreva2014time,smith2019quantizing}. It will also
be used operationally in Sec.~\ref{sec:winding_readout}: filtering the signal
frequency nonlocally selects the idler momentum at which the clock-cycle phase
is sampled.
 
Because the pair is generated by a nonlinear interaction, the global Hamiltonian
before pair creation is not simply \(H_C\otimes I+I\otimes H_S\). In the
undepleted-pump and narrowband limits, however, the generated single-pair state
satisfies the shifted Page--Wootters constraint
\begin{equation}
\label{eq:shifted_constraint}
\begin{aligned}
\hat J|\Psi\rangle\rangle&\simeq 0,
\\
\hat J
&=
H_C\otimes I
+
I\otimes H_S
-2\omega_p I .
\end{aligned}
\end{equation}
The constant term accounts for the pump energy in
Eq.~\eqref{eq:sfwm_energy_conservation}. The approximation in
Eq.~\eqref{eq:shifted_constraint} is controlled by the joint spectral bandwidth
of the source: for a cavity-limited microring source the fractional violation
scales as the loaded linewidth divided by the optical carrier frequency, of order
\(1/Q\). A derivation from the rotating-frame SFWM Hamiltonian is given in
Appendix~\ref{app:sfwm_paw_constraint}.

The signal and idler share one physical chain, so a single modulation waveform
acts on both bands, and the readout below requires that only the clock band
execute the topological cycle.  What separates them is the frequency dependence
of the tunable Mach--Zehnder couplers: the arm phases seen by two modes several
free spectral ranges apart differ by of order a radian
(Sec.~\ref{sec:protection_feasibility}), so one waveform traces two different
loops in the \((\delta,\Delta)\) plane.   The device is designed so that the idler
loop encircles the degeneracy while the signal loop does not.  If the separation
is imperfect and the signal band acquires a Chern number \(C_S\), the fringe
carries a contribution from the signal band as well; the combination that enters
the measured winding is worked out in Appendix~\ref{app:fringe_algebra}.
 
Conditioning the state \(|\Psi\rangle\rangle\) on a reading of the idler clock
gives the conditional signal state
\begin{equation}
\label{eq:conditional_signal_state}
|\psi_S(t)\rangle
=
{}_C\langle t|\Psi\rangle\rangle ,
\end{equation}
which obeys Schr\"odinger evolution under \(H_S\), up to a global pump-dependent
phase that cancels in the observables used below. The system photon therefore plays two roles: its conditional state evolves with respect to the clock reading, and it serves as the relational reference for the winding readout of Sec.~\ref{sec:winding_readout}.
 
\subsection{Covariant time states and directional clock readings}
\label{sec:covariant_clock_states}
 
The control parameters \(\delta\) and \(\Delta\) will be driven around a closed
loop, parametrised by a slow classical variable \(\tau\) with period \(T\), and
the topological content of that loop is the subject of
Sec.~\ref{sec:topological_cycle}.  Let \(\tau=0\) denote the reference point of
the loop, before it is applied.  The Page--Wootters time states are defined on the
lower band of this reference Hamiltonian. Writing
\(\omega_C(k_x)\equiv\omega_C(k_x;\tau=0)\) and
\(|u_C(k_x)\rangle\equiv|u_C(k_x;\tau=0)\rangle\), and passing to the continuum limit in the momentum label, the time states are
defined as
\begin{equation}
\label{eq:clock_time_state}
|t\rangle_C
=
\int_{-\pi}^{\pi} dk_x\;
w(k_x)\,
e^{-i\omega_C(k_x)t}\,
|u_C(k_x)\rangle ,
\end{equation}
where \(w(k_x)>0\) is a weight fixed below by the resolution of the identity.
 
The defining property of the family in Eq.~\eqref{eq:clock_time_state} is
covariance under the reference clock Hamiltonian,
\begin{equation}
\label{eq:clock_covariance}
e^{-iH_C s}|t\rangle_C=|t+s\rangle_C ,
\end{equation}
which holds for any dispersion and for any choice of \(w\). It follows because
\(e^{-iH_Cs}\) multiplies each momentum component by \(e^{-i\omega_C(k_x)s}\),
shifting the label \(t\) to \(t+s\).
 
Three properties of the family must be kept apart, and only the third requires an
approximation. The first is covariance, just established. The second is that the
readings resolve the identity. Integrating
Eq.~\eqref{eq:clock_time_state} over all readings and using
\(\int dt\, e^{-i[\omega_C(k_x)-\omega_C(k_x')]t}=2\pi\,\delta\!\left(\omega_C(k_x)-\omega_C(k_x')\right)\),
which on a monotonic branch has support only at \(k_x=k_x'\), gives
\begin{equation}
\label{eq:time_povm_resolution}
\begin{aligned}
\int_{-\infty}^{\infty}\!\! dt\;
&|t\rangle_C{}_C\langle t|
\\
&= 2\pi\!\int_{-\pi}^{\pi}\!\! dk_x\,
\frac{w^2(k_x)}{|\omega_C'(k_x)|}\,
|u_C(k_x)\rangle\langle u_C(k_x)| .
\end{aligned}
\end{equation}
The choice
\begin{equation}
\label{eq:povm_weight}
w(k_x)=\sqrt{\frac{|\omega_C'(k_x)|}{2\pi}}
\end{equation}
therefore makes the right-hand side a band projector, so that
\(\{|t\rangle_C{}_C\langle t|\}\) is a covariant positive-operator-valued
measure with respect to \(H_C\)~\cite{holevo1982probabilistic}, provided the
delta function in Eq.~\eqref{eq:time_povm_resolution} has support at a single
momentum.  That condition is examined below. The density-of-states weight in Eq.~\eqref{eq:povm_weight} is
constant wherever the dispersion is linear, in which case
Eq.~\eqref{eq:clock_time_state} reduces to the unweighted superposition of Bloch
states up to normalisation.
 
The third property is sharpness, and it is the one that is only approximate.
Distinct readings \(|t\rangle_C\) and \(|t'\rangle_C\) are close to orthogonal
only when the occupied momenta lie in a window over which \(\omega_C\) is
approximately linear. In that window the group velocity
\begin{equation}
\label{eq:group_velocity}
v_g(k_x)=a\,\frac{\partial\omega_C}{\partial k_x}
\end{equation}
is a single number, and it converts a displacement by one unit cell into the
calibratable time scale
\begin{equation}
\label{eq:t_cell_definition}
t_{\mathrm{cell}}=\frac{a}{|v_g|} .
\end{equation}
Covariance and the resolution of the identity are therefore exact, while the
timing resolution of the clock is set by the extent of the quasi-linear window.
 
For the Rice--Mele dispersion, Eq.~\eqref{eq:dispersion_closed_form} gives
\(\omega_C(k_x)=\omega_C(-k_x)\), so a frequency measurement alone selects
\(|k_x|\) and not the sign of \(k_x\). A state containing equal \(+k_x\) and
\(-k_x\) components has no directed clock propagation, and the family built on
the full zone does not resolve the identity, since the delta function in
Eq.~\eqref{eq:time_povm_resolution} then has support at two momenta. The operational clock is therefore obtained by directional excitation and
collection through a bus waveguide that phase-matches one travelling branch.
Restricting the integrals in Eqs.~\eqref{eq:clock_time_state}
and~\eqref{eq:time_povm_resolution} to that branch, written \(\mathcal B_C\) and
taken as \((0,\pi]\) for definiteness, the dispersion is strictly monotonic, the
delta function has support only at \(k_x=k_x'\), and the readings resolve the
identity on the branch.  The group velocity vanishes only at the branch
endpoints.

The Chern number and the factorisation of one pump cycle are properties of the closed band over the full \((k_x,\tau)\) torus, independent of which branch is populated in a given run. The directional branch fixes clock readings and momentum selection unambiguously, and the full winding is reconstructed by combining the two injection directions (Sec.~\ref{sec:winding_readout}).
 
Finally, \(\tau\) is not the Page--Wootters time.  It executes a closed loop in
the clock Hamiltonian, whereas the clock reading is the outcome label \(t\) of
the covariant family in Eq.~\eqref{eq:clock_time_state}.  The topological
invariant belongs to the cycle acting on the clock Hilbert space, while the
emergent time read from the clock is the relational variable with respect to
which the conditional system state evolves.


\section{Topological winding of the clock band}
\label{sec:topological_cycle}

What topological structure is carried by the clock band during one adiabatic
cycle?  The answer is the Wilson-loop winding of the lower clock band.  In the
adiabatic limit this winding has an operator-level expression on the clock
Hilbert space: the cycle holonomy factorises into a clock-band lattice
translation by \(C\) unit cells, a momentum-periodic geometric factor, and a
dynamical phase.  This section establishes the Chern number, the associated
Wilson-loop winding, and the representation of that winding targeted by the
relational readout of Sec.~\ref{sec:winding_readout}.

\subsection{Clock-band Chern number}
\label{sec:cycle_chern}

The gap of Eq.~\eqref{eq:rice_mele_hamiltonian} is
\(2\sqrt{d_x^2+d_y^2+d_z^2}\) and closes only where \(\mathbf d=0\).  Setting
\(d_y=0\) forces \(\sin k_x=0\); substituting \(k_x=\pi\) into \(d_x=0\) gives
\(2J_x\delta=0\), and \(d_z=0\) gives \(\Delta=0\).  The unique gap-closing
point is therefore
\begin{equation}
\label{eq:dirac_point}
(\delta,\Delta,k_x)=(0,0,\pi).
\end{equation}
The two control knobs are cycled adiabatically around it,
\begin{equation}
\label{eq:pump_cycle}
\delta(\tau)=R\cos\frac{2\pi\tau}{T},
\qquad
\Delta(\tau)=R\,J_x\sin\frac{2\pi\tau}{T},
\end{equation}
with \(0<R<1\) dimensionless.  The loop encircles the degeneracy without
touching it, and Eq.~\eqref{eq:dispersion_closed_form} gives the minimum gap on
the cycle as \(2RJ_x\), attained at the zone edge.

The topological content of the cycle is the Chern number of the lower clock-band
bundle over the \((k_x,\tau)\) torus,
\begin{equation}
\label{eq:chern_number}
\begin{aligned}
C&=\frac{1}{4\pi}\int_0^T\!\!\int_{-\pi}^{\pi}
\hat{\mathbf n}\cdot
\left(\partial_\tau\hat{\mathbf n}\times\partial_{k_x}\hat{\mathbf n}\right)
dk_x\,d\tau ,
\\[4pt]
\hat{\mathbf n}&=\frac{\mathbf d}{|\mathbf d|} .
\end{aligned}
\end{equation}
The orientation is fixed so that positive \(C\) corresponds to transport in the
\(+x\) direction in the usual Thouless-pump convention.  With this convention,
the cycle of Eq.~\eqref{eq:pump_cycle} carries
\begin{equation}
\label{eq:chern_value}
C=+1 .
\end{equation}
Numerical checks in Appendix~\ref{app:numerics} verify this orientation and the
corresponding transport.

For reference, the same integer appears in the standard filled-band formulation
of Thouless pumping.  A state that samples the Brillouin zone uniformly has
Wannier-centre displacement
\begin{equation}
\label{eq:thouless_transport}
\Delta X_W
=
\frac{1}{2\pi}\int_0^T\!\!\int_{-\pi}^{\pi}
\Omega(k_x,\tau)\,dk_x\,d\tau
=
C\,a ,
\end{equation}
with \(\Omega\) the Berry curvature in the orientation of
Eq.~\eqref{eq:chern_number}.  This is the standard Wilson-loop, or hybrid-Wannier,
form of quantised pumping; below it is used as the topology of the clock band
rather than as the observable measured by a packet displacement.

\begin{figure}[t]
\centering
\includegraphics[width=\columnwidth]{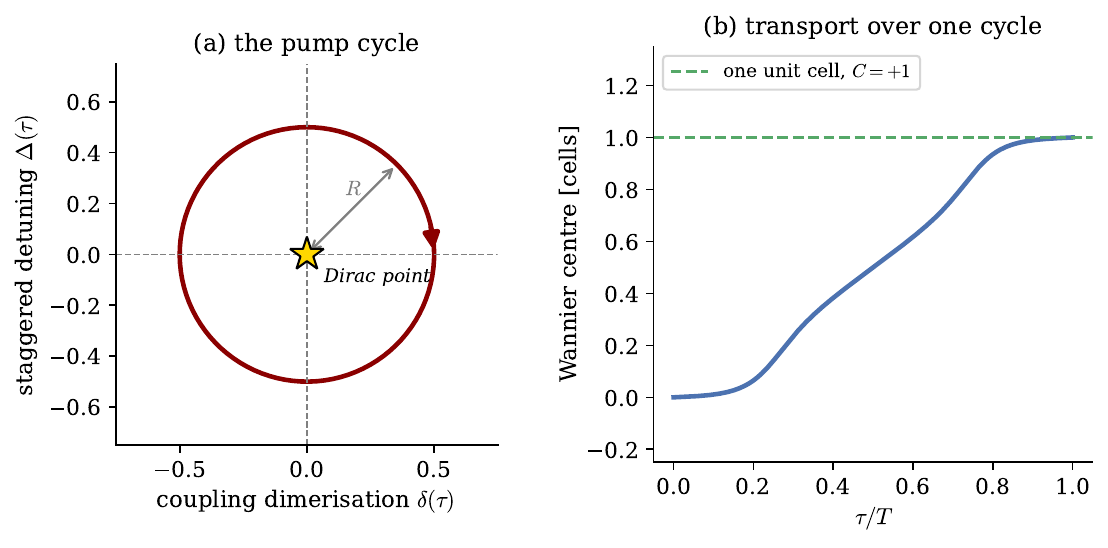}
\caption{%
(a) The cycle of Eq.~\eqref{eq:pump_cycle} in the \((\delta,\Delta)\) plane,
a circle of radius \(R\) encircling the Dirac point without touching it; the
minimum gap on the loop is \(2RJ_x\).  (b) Hybrid Wannier centre of the lower
band through one cycle, computed from the King-Smith--Vanderbilt polarisation.
It advances by one unit cell in the \(+x\) convention, corresponding to
Eq.~\eqref{eq:thouless_transport} with \(C=+1\).}
\label{fig:cycle}
\end{figure}

\subsection{Wilson-loop winding and clock-band holonomy}
\label{sec:factorisation}

At fixed crystal momentum \(k_x\), the pump cycle defines a closed loop in the
parameter plane.  In the adiabatic limit, \(T\gg 1/(2RJ_x)\), a lower-band state
accumulates a dynamical phase and the geometric Berry phase
\begin{equation}
\label{eq:berry_phase}
\begin{aligned}
\gamma(k_x)&=\oint A_\tau(k_x,\tau)\,d\tau ,
\\[4pt]
A_\tau&=i\langle u(k_x,\tau)|\partial_\tau u(k_x,\tau)\rangle .
\end{aligned}
\end{equation}
Because the parameter loop is closed, \(\gamma(k_x)\) is gauge invariant modulo
\(2\pi\) at each momentum fibre.  Its unwrapped change across the Brillouin zone
is fixed by the Chern number.  Applying Stokes' theorem on the cylinder
\(k_x\in[-\pi,\pi]\), \(\tau\in[0,T]\), in a gauge periodic in \(\tau\) and
smooth in \(k_x\), gives
\begin{equation}
\label{eq:winding_relation}
\gamma(\pi)-\gamma(-\pi)=-2\pi C .
\end{equation}
When \(C\neq 0\), no gauge can be smooth and periodic on the whole torus; this
obstruction is precisely the Wilson-loop winding.

The same winding has a useful operator expression on the clock band.  Splitting
off the linear part, Eq.~\eqref{eq:winding_relation} allows
\begin{equation}
\label{eq:gamma_split}
\gamma(k_x) = -Ck_x + \gamma_{\mathrm{per}}(k_x),
\end{equation}
where \(\gamma_{\mathrm{per}}\) is periodic on the Brillouin zone.  Since
\(k_x\) is dimensionless, multiplication by \(e^{-iCk_x}\) in the momentum
representation is the lattice translation \(\hat T_{Ca}\) by \(+Ca\) in real
space.  The adiabatic cycle holonomy on the
lower clock band therefore factorises as
\begin{equation}
\label{eq:factorisation}
\hat U_{\mathrm{cycle}}
=
\hat T_{Ca}\,
e^{i\gamma_{\mathrm{per}}(\hat k_x)}\,
e^{-i\int_0^T \omega_C(\hat k_x,\tau)\,d\tau} .
\end{equation}
Equation~\eqref{eq:factorisation} is the operator-level consequence of the
Wilson-loop winding used by the readout protocol below.  It is exact in the
adiabatic limit, diagonal in \(k_x\), and does not assume a uniform momentum
weight or a particular input state.  The last factor is the dynamical phase
fibre,
\(\phi_{\mathrm{dyn}}(k_x)=-\int_0^T\omega_C(k_x,\tau)\,d\tau\).

\begin{figure}[t]
\centering
\includegraphics[width=\columnwidth]{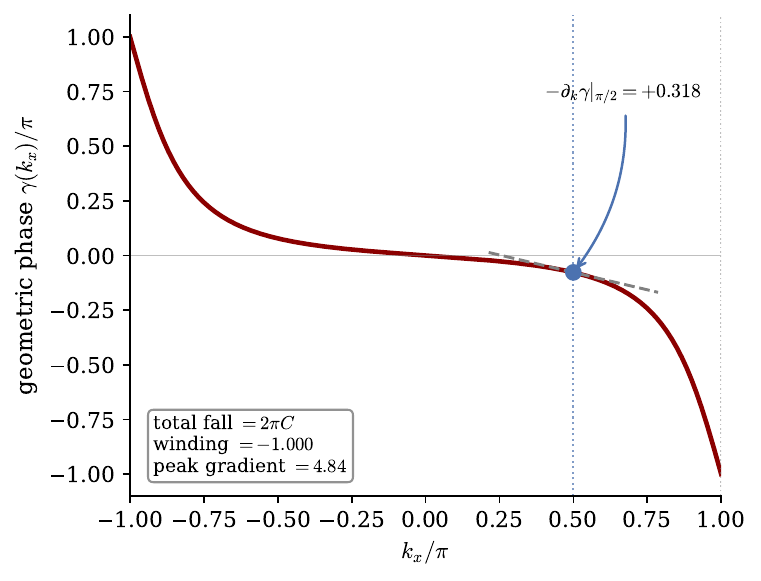}
\caption{%
Geometric phase \(\gamma(k_x)\) accumulated over one pump cycle, computed as a
Wilson loop in \(\tau\) at each momentum.  The unwrapped profile falls by
\(-1.000\) turns across the zone, which is Eq.~\eqref{eq:winding_relation} with
\(C=+1\).  The dashed line is the tangent at the operating point
\(k_x=\pi/2\), where the local slope \(-\partial_{k_x}\gamma=0.318\) lies well
below the zone average of unity; the gradient peaks at \(4.84\) near the zone
edge, which is what sets the sampling condition of
Eq.~\eqref{eq:sampling_condition}.  This profile is the quantity the fringe of
Sec.~\ref{sec:winding_readout} samples.
}
\label{fig:gamma}
\end{figure}

The quantised content of the holonomy is the translation factor itself: the
cycle contains a state-independent shift of the clock-band lattice by \(C\) unit
cells.  The remaining momentum-dependent phases determine how any particular
wavepacket moves in real space.  Thus the topological object is not a photon
delay or a calibrated tick duration, but the winding of the clock-band holonomy.

\subsection{What the readout targets}
\label{sec:readout_target}

The relational measurement of Sec.~\ref{sec:winding_readout} targets the winding
of \(\gamma(k_x)\) (Fig.~\ref{fig:gamma}), not the real-space displacement of a
heralded photon packet. This distinction is important because the source prepares a narrow momentum
window, whereas the Thouless displacement in Eq.~\eqref{eq:thouless_transport}
is a full-zone quantity.  A packet measurement would sample the local gradient
of the accumulated phase at the occupied momenta; the invariant is recovered
instead by scanning the phase across the Brillouin zone and unwrapping its
winding. The wavepacket-displacement calculation, including the numerical value for the
heralded operating point, is given in
Appendix~\ref{app:packet_displacement} and Fig.~\ref{fig:packet}.


%
%

\section{The interferometric winding readout}
\label{sec:winding_readout}
A photon that traverses the cycle accumulates both phases of
Eq.~\eqref{eq:factorisation}, and a detector sees only their sum.  Extracting the
winding therefore requires separating the two, moving the heralded momentum
across the zone, and supplying a reference against which a phase is defined at
all.  The entangled partner provides the last two at once.

\begin{figure*}[t]
\centering
\includegraphics[width=\textwidth]{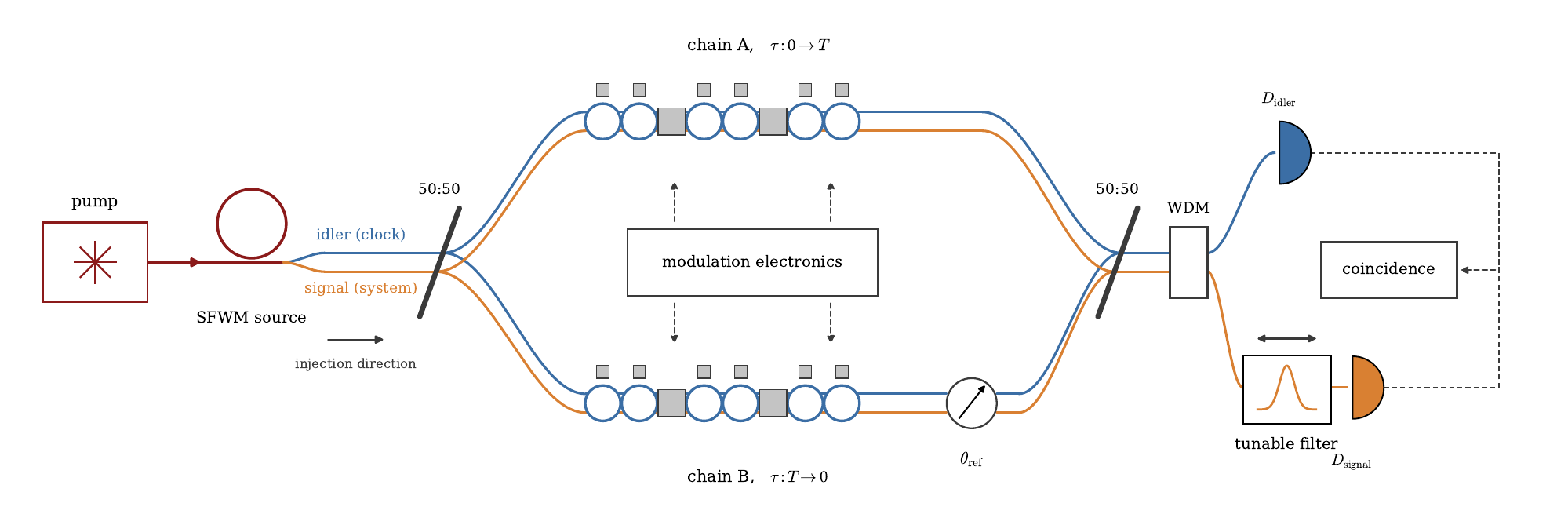}
\caption{%
The winding readout in the parallel implementation of
Sec.~\ref{sec:implementation}.  A degenerate pump drives spontaneous four-wave
mixing in a microring source, and the pair is split into an idler, which
carries the clock, and a signal, which serves as the system and as the phase
reference.  The idler amplitude is divided between two nominally identical
Rice--Mele chains, which the modulation electronics drive through the loop of
Eq.~\eqref{eq:pump_cycle} in opposite orientations, \(\tau:0\to T\) in one and
\(\tau:T\to0\) in the other, so that the dynamical phase cancels and the fringe
difference isolates \(2\gamma(k_x)\).  A tunable narrowband filter on the
signal steps the heralded idler momentum through the anticorrelation of
Eq.~\eqref{eq:energy_anticorrelation}, without touching the clock.  Unwrapping
the coincidence fringe across the zone returns the integer of
Eq.~\eqref{eq:winding_readout}.
}
\label{fig:setup}
\end{figure*}

The translation factor of Eq.~\eqref{eq:factorisation} is the linear part of
\(\gamma\), so a measurement of \(\gamma\) reaches it. Traversing the loop of
Eq.~\eqref{eq:pump_cycle} backward, \(\tau:T\to 0\), reverses the orientation of
the parameter loop, so \(\gamma\to-\gamma\). The phase accumulated by a
lower-band state at fixed \(k_x\) is accordingly
\begin{equation}
\label{eq:fwd_bwd_phases}
\begin{aligned}
\Phi_{\mathrm{fwd}}(k_x)&=\phi_{\mathrm{dyn}}(k_x)+\gamma(k_x),
\\[4pt]
\Phi_{\mathrm{bwd}}(k_x)&=\phi_{\mathrm{dyn}}(k_x)-\gamma(k_x),
\end{aligned}
\end{equation}
and the difference isolates twice the geometric phase with the dynamical
contribution cancelled identically,
\begin{equation}
\label{eq:fringe_difference}
\begin{aligned}
\Delta\Phi(k_x)
&=
\Phi_{\mathrm{fwd}}(k_x)-\Phi_{\mathrm{bwd}}(k_x)
\\[4pt]
&=
2\,\gamma(k_x).
\end{aligned}
\end{equation}
The cancellation in Eq.~\eqref{eq:fringe_difference} is not a small-parameter
statement. It holds because the backward traversal visits the same parameter
values on the same schedule in reverse, so the dynamical integral is unchanged
while the orientation of the loop, and hence the sign of \(\gamma\), is
reversed. Operationally the two traversals are superposed in a two-photon
coincidence interferometer: the pair amplitude that experienced the forward cycle
is brought into interference with the amplitude that experienced the backward
cycle, and the coincidence rate between the signal and idler detectors oscillates
in the relative phase, which is read out as \(\Delta\Phi(k_x)\) at the heralded
momentum. The fringe algebra, including the role of the signal photon as the
reference arm, is given in Appendix~\ref{app:fringe_algebra}; how the two traversals could be realised is discussed in Sec.~\ref{sec:implementation}.

The fringe gives \(\gamma\) at one momentum, while the winding needs the whole
zone, and the heralded idler occupies a window of \(\Delta k_x\approx 0.041\)
rad at the budgeted quality factor (Sec.~\ref{sec:sampling}). The scan is
therefore performed on the partner. Energy anticorrelation means a narrowband
filter on the signal at \(\omega_s\) heralds the idler at
\(2\omega_p-\omega_s\), which the dispersion, strictly monotonic on a single
directional branch, turns into a definite \(k_x\)
(Sec.~\ref{sec:momentum_sign}). Stepping the filter steps the heralded momentum, and the clock
photon is never touched.

The signal photon settles the reference as well. An interferometric phase exists only
against something, and an external optical reference would reintroduce the very
time standard the construction removes. Pump phase noise and common path drifts
enter both members of a pair equally and cancel in the coincidence, leaving the
differential fringe alone.

Collecting the fringe phase at \(N_{\mathrm{scan}}\) settings across the zone and
unwrapping it reconstructs \(2\gamma(k_x)\) up to a constant, whose accumulated
change is fixed by Eq.~\eqref{eq:winding_relation},
\begin{equation}
\label{eq:winding_readout}
W\equiv\frac{1}{2\pi}\Big[\Delta\Phi\Big]_{k_x=-\pi}^{k_x=+\pi}=-2C .
\end{equation}
No fractional value can appear, since the winding of a phase around a closed
loop is an integer by construction. The sign of \(W\) follows from which
traversal is labelled forward, and deliberately relabelling the two is a
consistency check.

\subsection{Resolving the sign of the momentum}
\label{sec:momentum_sign}
That proviso is not a technicality. Since \(\omega_C(k_x)=\omega_C(-k_x)\), a
frequency filter heralds \(|k_x|\) rather than \(k_x\), and if both momenta are
populated the two contributions enter the fringe with nearly opposite phases.
The fringe does not vanish, which is what makes the failure dangerous: with
equal population the visibility stays above \(0.8\) across most of the zone
while the measured phase is identically zero, so the scan returns a flat profile
and \(W=0\).

The resolution is the directional bus injection already required for the clock
family to resolve the identity, Sec.~\ref{sec:covariant_clock_states}. Exciting
the chain through a bus waveguide that phase-matches one propagation direction
populates a single half zone, and the scan then proceeds over \(k_x\in(0,\pi]\)
with each filter setting heralding a single momentum.

The integer is then recovered from two such half-zone records. Reversing the injection direction in a second run populates the complementary
half zone. Concatenating the two half-zone fringe records, with the samples of
the second run entering in reversed \(k_x\) order, reconstructs the fringe over
the full zone, and its unwrapped winding is \(-2C\) by
Eq.~\eqref{eq:winding_readout}, with no symmetry assumption anywhere. The cost is
a second data set and one relative phase between the runs. That relative phase is
a single constant; because the winding depends only on phase differences around
the closed scan, it drops out and need not be known.

The constant enters the closed sum through the
two junction steps where the half-zone records meet, with opposite signs. If it
happens to sit near \(\pi\), one junction difference can wrap while the other does
not, shifting the estimate by one turn. The protocol detects this automatically:
the true winding \(-2C\) is even, a single junction wrap error makes the estimate
odd, and the parity of the result therefore serves as a built-in consistency
flag. It is cleared by re-referencing the second run, that is, by shifting its
record by a trial constant chosen to minimise the junction steps, and
unwrapping again.

The clean Rice--Mele cycle obeys \(d_x(-k_x)=d_x(k_x)\),
\(d_y(-k_x)=-d_y(k_x)\), and \(d_z\) independent of \(k_x\), from which the
geometric phase inherits the parity
\begin{equation}
\label{eq:gamma_parity}
\gamma(-k_x)=-\gamma(k_x)
\qquad (\mathrm{mod}\ 2\pi),
\end{equation}
verified numerically to \(3.6\times10^{-15}\). On such a device one run
suffices: the half-zone winding is exactly half the full one, and a single
directional run returns \(C\) through \(W_{\mathrm{half}}=-C\) at half the
acquisition time. That shortcut rests on a symmetry of the clean model rather
than on the gap, and disorder degrades it continuously, so the two-run stitch is
the primary protocol here.

%
%
%
%
%
%
%

\section{Protection}
\label{sec:protection_feasibility}

The readout of Sec.~\ref{sec:winding_readout} is protected in two distinct
senses. The gap pins the ideal winding against deformations of the device and of
the pump loop, so the integer cannot move unless the gap closes somewhere on the
Bloch torus. The estimator that reads it is digital, so below a threshold set by
the sampling margin the probability of returning the wrong value falls
exponentially. Neither layer is unconditional, and both are stated below as
inequalities with their conditions attached.

\begin{figure*}[t]
\centering
\includegraphics[width=\textwidth]{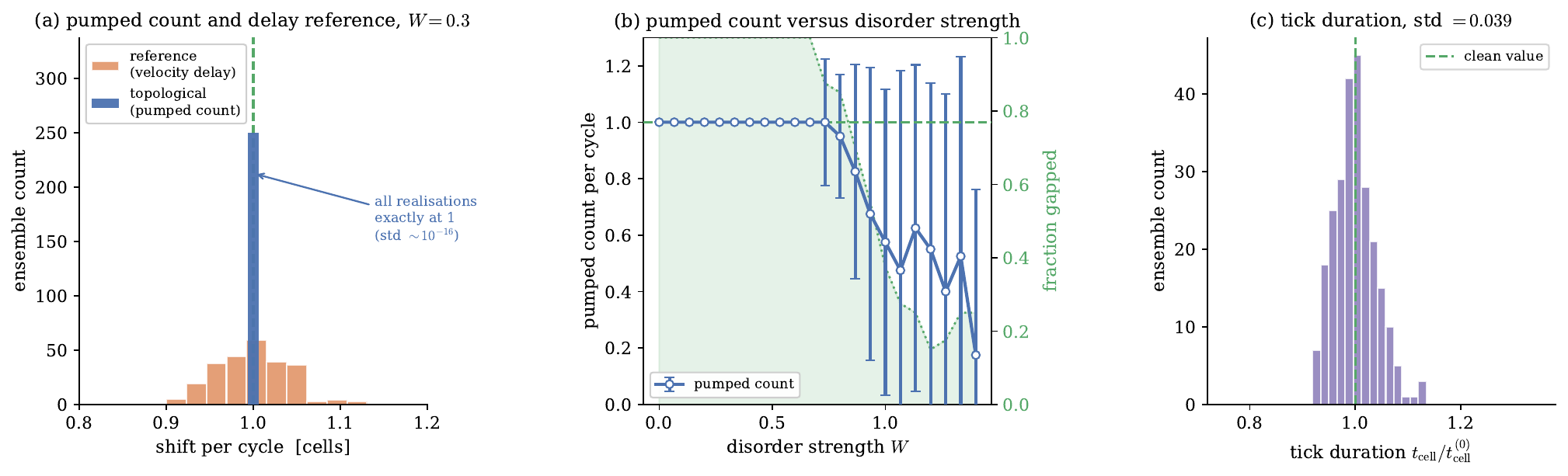}
\caption{%
Quenched disorder at strength \(W\), from Resta polarisation transport on a
finite ring. (a) At \(W=0.3\) the pumped count is pinned at one cell across the
whole ensemble, with a standard deviation at the level of machine precision,
while a group-velocity delay reference drifts with a spread set linearly by the
disorder. (b) The count stays quantised until the disorder closes the bulk gap;
the shaded curve is the fraction of realisations that remain gapped. (c) The
tick duration inherits the disorder of the group velocity and spreads by
\(3.9\%\) over the same ensemble. The count is protected; the scale is not.
}
\label{fig:disorder}
\end{figure*}

\subsection{Gap protection of the winding}
\label{sec:gap_protection}

The protected object of this paper is the winding of \(\gamma(k_x)\), and its
protection is demonstrated directly. An ensemble of random smooth deformations of
the Bloch vector field,
\begin{equation}
\label{eq:deformation_ensemble}
\mathbf d(k_x,\tau)\;\longrightarrow\;
\mathbf d(k_x,\tau)+W J_x\,\mathbf f(k_x,\tau) ,
\end{equation}
with each Cartesian component of \(\mathbf f\) an independent random truncated
Fourier series on the torus, periodic and of unit root-mean-square amplitude, is
applied to the clean cycle. For each realisation the winding of the deformed
\(\gamma(k_x)\) is computed together with the minimum gap encountered anywhere on
the torus at any point along the deformation path.

The result is a clean dichotomy. Every realisation whose gap stayed open along
the entire path has winding pinned at the clean integer to machine precision, and
every realisation whose winding differs from the clean value experienced a gap
closing somewhere en route, after which the integer reopens at a different value.
The winding moves only through a closing, never through a drift. This is the
operative content of gap protection for the observable the experiment measures,
and the deformation family is the right one for it: the winding of \(\gamma(k_x)\)
is defined on the Bloch torus and is the appropriate invariant for perturbations
that preserve crystal momentum, such as the coupler-induced loop distortions of
Sec.~\ref{sec:sampling} and slow parameter drifts.

Quenched site-by-site disorder is a different perturbation class, since it breaks
translation invariance and takes the analysis outside the Bloch description.
There the corresponding statement is the Niu--Thouless result that the
transported charge of a filled band remains quantised in the presence of disorder
and interactions while the relevant gap survives~\cite{niu1984quantised}. A
real-space simulation confirms it for this platform: the pumped count of the
lower band, computed by Resta polarisation transport~\cite{resta1998quantum} on
disordered rings, stays pinned at one cell with machine-precision rigidity until
the disorder closes the bulk gap. The two statements are complementary faces of
the same protection, one in Bloch space and one in real space, and both are
recorded in Appendix~\ref{app:numerics}.

That quantised pumping survives disorder until the gap closes is consistent with
its direct observation in laser-written photonic
lattices~\cite{cerjan2020thouless}. Nonlinearity is likewise not a threat to the
integer: nonlinear Thouless pumping remains quantised through a distinct
mechanism~\cite{jurgensen2021quantized}, whereas here the \(\chi^{(3)}\)
interaction shapes the joint spectrum rather than entering the constraint
(Appendix~\ref{app:sfwm_paw_constraint}).

A chiral-symmetry-breaking term sharpens what kind of protection this is. Adding
a staggered contribution \(\lambda\,\sigma_z\) that is held fixed, and is not part
of the cycle, leaves the winding exactly at its clean value until \(\lambda\)
closes the gap near \(\lambda\approx 0.5\,J_x\), beyond which the invariant is no
longer defined. The protection is therefore by the gap alone, which is strictly
stronger than protection by a symmetry: an invariant defined through a symmetry
drifts continuously once that symmetry is broken, whereas the winding cannot move
at all while the gap is open. This distinction also grades the two half-zone
routes of Sec.~\ref{sec:momentum_sign}. The parity shortcut borrows
Eq.~\eqref{eq:gamma_parity}, a property of the clean model, and is exposed when
that parity is broken; the full-zone winding of the two-run stitch relies only on
the gap. The term \(\lambda\,\sigma_z\) is not itself a test of the shortcut,
since it leaves \(d_z\) independent of \(k_x\) and therefore preserves the
specific parity the shortcut uses. That test is contained in the deformation
ensemble of Eq.~\eqref{eq:deformation_ensemble}, whose random fields generically
break the parity: under those deformations the premise of the shortcut fails
while the full-zone winding remains pinned as long as the gap stays open.

Finally, the same numerics show what is not protected. The tick duration
\(t_{\mathrm{cell}}=a/|v_g|\) of Eq.~\eqref{eq:t_cell_definition} inherits the
disorder of the group velocity and spreads by several percent over the same
ensemble (Fig.~\ref{fig:disorder}). The protected content is the integer,
whether expressed as the transported count or as the fringe winding; the
conversion of that integer into a duration is a calibratable, device-dependent
scale, in the same way that the SI second fixes a count rather than a duration.
Two nominally identical devices, differing only in fabrication disorder, will
therefore have slightly different tick durations, and a calibration of
\(t_{\mathrm{cell}}\) performed on one is not transferable to another. What is
transferable is the integer, and the measurement of
Sec.~\ref{sec:winding_readout} never passes through \(t_{\mathrm{cell}}\) at all.

\subsection{Sampling and filter requirements}
\label{sec:sampling}

Unwrapping succeeds when adjacent samples of the fringe differ by less than
\(\pi\). With samples spaced by \(\delta k\) in the heralded momentum, the
requirement is
\begin{equation}
\label{eq:sampling_condition}
\delta k \max_{k_x}\left|\partial_{k_x}\Delta\Phi\right|
= 2\,\delta k \max_{k_x}\left|\partial_{k_x}\gamma\right| < \pi .
\end{equation}
The gradient of the geometric phase is largest where the Berry curvature of the
cycle concentrates, near the zone edge \(k_x=\pi\), and at \(R=0.5\) its computed
peak is \(\max_{k_x}|\partial_{k_x}\gamma|=4.84\) rad per rad
(Fig.~\ref{fig:gamma}), so that
Eq.~\eqref{eq:sampling_condition} gives \(\delta k<0.32\) rad and a minimum of
\(N_{\mathrm{scan}}>4\max_{k_x}|\partial_{k_x}\gamma|\approx 20\) filter settings
across the full zone. Noise margin favours denser sampling, and the budget of
Sec.~\ref{sec:feasibility_budget} adopts \(N_{\mathrm{scan}}=64\), at which the
largest ideal fringe step is \(\Delta_{\max}=0.95\) rad, inside the unwrapping
window with margin \(\pi-\Delta_{\max}=2.19\) rad. This condition is set by
\(\gamma(k_x)\), which is a Wilson loop over the whole cycle, and is therefore
independent of where on the cycle the dispersion is evaluated.

The corresponding demand on the signal filter is not, since it runs through the
dispersion: a momentum step \(\delta k\) maps to a frequency step
\(\delta f=|v_g(k_x)|\,\delta k/(2\pi a)\). The clock states of
Eq.~\eqref{eq:clock_time_state} are built on the reference Hamiltonian at
\(\tau=0\), which by Eq.~\eqref{eq:pump_cycle} is \((\delta,\Delta)=(R,0)\), and
there \(\partial\omega_C/\partial k_x=0.474\,J_x\) at the band centre. With
\(J_x/2\pi=4\) GHz and \(\delta k=2\pi/64\) this fixes the step spacing at
\(\delta f\approx 0.19\) GHz at the band centre. The spacing is not uniform in
frequency, since the group velocity vanishes at the zone edge: the lower band
spans \(J_x\), or \(4\) GHz, across the half zone, so \(32\) settings per half
zone correspond to a mean spacing near \(125\) MHz. Both figures sit within the
resolution of standard tunable filters and small against the free spectral range.

The filter passband, as distinct from its step spacing, sets a separate and more
delicate requirement, and the band edge is where it binds. By
Eq.~\eqref{eq:dispersion_closed_form} the group velocity vanishes at
\(k_x=\pi\), so the frequency-to-momentum map is quadratic there,
\(\omega_C(k_x)\approx\omega_C(\pi)+\alpha(k_x-\pi)^2\) with \(\alpha\) of order
\(J_x/2\). A passband of width \(\Delta\omega\) therefore heralds a momentum
window of width \(\sim\sqrt{\Delta\omega/\alpha}\) at the edge, and even the bare
ring linewidth heralds a window of order \(0.2\) rad there. At the band centre
the same linewidth heralds \(\Delta k_x\approx 0.041\) rad, about
\(0.7\%\) of the zone.

The measured fringe phase at such a setting is not the point value
\(2\gamma(k_j)\) but the phase of the windowed phasor
\begin{equation}
\label{eq:windowed_phasor}
Z_j=\int_{H_j} dk_x\, W_j(k_x)\, e^{i2\gamma(k_x)} ,
\qquad
\varphi_j=\arg Z_j ,
\end{equation}
with visibility \(V_j=|Z_j|/\int_{H_j} dk_x\,W_j(k_x)\), where \(W_j\) is the
passband lineshape mapped through the clock dispersion and \(H_j\) is the half
zone populated by the directional injection of Sec.~\ref{sec:momentum_sign}.
Smoothing of this kind cannot change a winding without dragging the phasor
through zero, so the windowed fringe continues to carry the integer as long as
every \(V_j\) stays finite and every wrapped step stays below \(\pi\). Two
features of the corrected operating point bear on this. The danger is confined to
the junction sample at \(k_x=\pi\), where the window is widest and the two
half-zone records are stitched together; and since the mean step spacing of
\(125\) MHz is comparable to the passband, adjacent windows overlap, so the
samples are correlated rather than independent. Overlap of this kind is
oversampling and does not by itself move a winding, but it does enter the
windowed phasor, and the check of Appendix~\ref{app:numerics} is performed at the
reference dispersion for that reason. A window at the ring linewidth returns
\(-2C\) for Gaussian and Lorentzian lineshapes alike, with worst-sample
visibilities of \(0.83\) and \(0.74\); a \(0.15\) GHz passband still
returns the integer, while a \(0.3\) GHz Lorentzian passband misreads the
junction step and shifts the estimate by one turn, an odd result that the parity
flag of Sec.~\ref{sec:momentum_sign} detects. The passband must therefore
not exceed approximately \(0.15\) GHz, and should preferably be apodised, since
Lorentzian tails smear more phase into the window at equal nominal width.

One further systematic is inherited from the tunable couplers. Their arm phases
vary with optical frequency, so the loop traced in the \((\delta,\Delta)\) plane
varies slightly across the clock band and not only between the clock and system
bands. This deforms \(\gamma(k_x)\) smoothly, and it is survivable for exactly
the reason the observable was chosen: a winding is discrete and cannot respond to
a smooth \(k_x\)-dependent deformation of the loop family unless the gap closes
for some member of it, which the budgeted parameters keep far away.

\subsection{Digital protection: the threshold inequality}
\label{sec:digital_protection}

The metrologically decisive property of the winding readout is that the estimate
it returns is an integer. Explicitly, the protocol returns
\begin{equation}
\label{eq:winding_estimator}
\widehat W=\mathrm{round}\!\left[
\frac{1}{2\pi}\sum_{j=1}^{N_{\mathrm{scan}}}
\mathrm{wrap}\!\left(\varphi_{j+1}-\varphi_j\right)\right] ,
\end{equation}
where \(\varphi_j\) is the measured fringe phase at the \(j\)th filter setting,
the sum runs once around the closed scan, and \(\mathrm{wrap}\) reduces each
difference to \((-\pi,\pi]\). A single step is misread only if its noise exceeds
the margin \(\pi-|\Delta_j|\) left by the ideal step \(\Delta_j\). For
independent Gaussian fringe noise of standard deviation \(\sigma_\varphi\) per
sample, adjacent differences carry noise \(\sqrt2\,\sigma_\varphi\), and a union
bound over the scan gives
\begin{equation}
\label{eq:threshold_bound}
P\!\left(\widehat W\neq W\right)
\le
N_{\mathrm{scan}}\,
\mathrm{erfc}\!\left(\frac{\pi-\Delta_{\max}}{2\sigma_\varphi}\right).
\end{equation}
Equation~\eqref{eq:threshold_bound} is an inequality with its conditions
attached. The readout is not noise free; it is threshold protected, with the
misidentification probability falling exponentially in \(1/\sigma_\varphi^2\)
below threshold.

Monte Carlo simulation of the full estimator on the computed fringe profile
confirms the bound. At \(N_{\mathrm{scan}}=64\) and \(\sigma_\varphi=0.4\) rad,
no error occurred in 6000 trials against a bound of \(7\times10^{-3}\); at
\(\sigma_\varphi=0.7\) rad the observed error rate is \(0.16\), the threshold
having been crossed and the bound having become vacuous
(Fig.~\ref{fig:winding}). For contrast, an analog reference built on the packet
displacement of Appendix~\ref{app:packet_displacement} degrades linearly in the
same noise, with no threshold and no rounding to fall back on. This
exponential-versus-linear separation below threshold is the concrete metrological
content of carrying the information in an integer, and it requires no hardware
beyond the interferometer already needed for the readout.

\begin{figure}[t]
\centering
\includegraphics[width=\columnwidth]{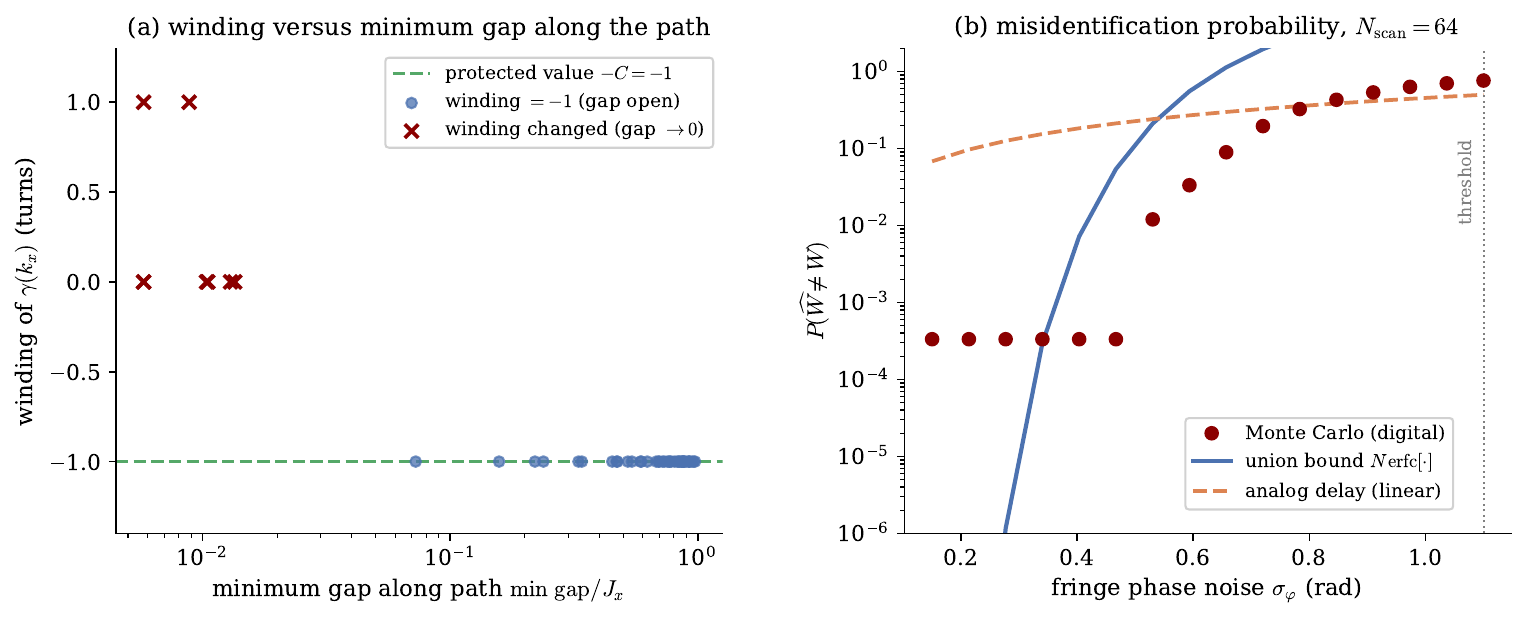}
\caption{%
(a) Winding of the deformed \(\gamma(k_x)\) against the smallest gap encountered
along the deformation path of Eq.~\eqref{eq:deformation_ensemble}. Every
realisation that stayed gapped returns the clean value; every realisation that
returns a different integer passed through a closing. The winding moves by
jumping, never by drifting.
(b) Probability of returning the wrong integer against fringe phase noise, at
\(N_{\mathrm{scan}}=64\). Points are Monte Carlo on the computed fringe profile,
the solid line is the union bound of Eq.~\eqref{eq:threshold_bound}, and the
dashed line is an analog delay estimate degrading linearly in the same noise.
The two separate below threshold.
}
\label{fig:winding}
\end{figure}

Two clarifications complete the statement. First, \(\Delta_{\max}\) in
Eq.~\eqref{eq:threshold_bound} denotes the largest noiseless wrapped step of the
phases actually fed to the estimator. For point samples of \(2\gamma(k_x)\) this
is \(0.95\) rad. For the windowed phases of Eq.~\eqref{eq:windowed_phasor} the
raw junction step is larger, up to \(2.5\) rad at the \(0.15\) GHz passband, but the junction bias is deterministic and the re-referencing step of
Sec.~\ref{sec:momentum_sign} minimises it: after re-referencing, the operative
step is \(0.94\) to \(1.26\) rad depending on lineshape and passband, and Monte
Carlo on the re-referenced windowed phases at \(\sigma_\varphi=0.4\) rad returns
error rates between \(2\times10^{-4}\) and \(8\times10^{-4}\), inside the bound. Second, the reduced fringe visibility enters through \(\sigma_\varphi\)
itself rather than through the protection. Shot-noise-limited phase extraction
scales as \(\sigma_\varphi\simeq 1/(V\sqrt{N_{cc}})\) with \(N_{cc}\) the
coincidence count per setting, so the worst-sample visibility raises the count
needed per setting by \(1/V^2\approx 2.6\). This is a rate cost, not a protection
cost.


\section{Feasibility}
\label{sec:feasibility}

\subsection{Implementing the two traversals}
\label{sec:implementation}

The fringe of Eq.~\eqref{eq:fringe_difference} requires the forward and backward
traversals to be superposed coherently, and spontaneous four-wave mixing supplies
no trigger: pairs are born at random times within a continuous-wave drive.

A time-bin realisation in the manner of Franson~\cite{franson1989bell} would be
the natural first choice, alternating forward and backward cycles in consecutive
bins of duration \(T\) and post-selecting the central coincidence peak. The
filtering that the momentum scan requires excludes it. A passband of \(0.15\) GHz
gives the filtered signal photon an arrival-time spread of two to three
nanoseconds, against a bin separation of \(T\approx 0.8\) ns, so the side bins
cannot be resolved and the post-selection fails independently of detector jitter.
The two requirements are not simultaneously satisfiable: narrowing the passband
to sharpen the momentum herald widens the arrival-time distribution in the same
proportion.

The two traversals are therefore run in parallel. The pair amplitude is split
between two nominally identical chains driven through opposite loop orientations,
so that both traversals proceed simultaneously and the lifetime requirement is
\(T<\tau_{\mathrm{ph}}\) rather than \(2T<\tau_{\mathrm{ph}}\). The price is that
the dynamical phases of the two arms no longer cancel identically, since
fabrication disorder makes the two dispersions differ. The residual
\(\delta\phi_{\mathrm{dyn}}(k_x)\) is, however, a smooth single-valued function
of \(k_x\), periodic across the zone, and therefore contributes exactly zero
winding to the fringe. It costs only unwrapping margin, through its gradient, of
order \(\delta J_x T\approx 0.2\) rad per radian of \(k_x\) for percent-level
coupling mismatch, small against the \(2.19\) rad margin of
Sec.~\ref{sec:sampling}; a constant carrier-offset mismatch between the chains
drops out entirely. The winding readout thus tolerates a mismatched
interferometer that would ruin a phase-value measurement, which is one more
expression of the count-versus-scale distinction. Appendix~\ref{app:fringe_algebra}
gives the fringe algebra for two physically distinct chains, where the signal
propagation phases are no longer identical by construction.

Running in parallel fixes the device scale. A heralded momentum width of
\(\Delta k_x\approx 0.041\) rad corresponds to a packet extended over roughly
\(25\) unit cells, and at \(\partial\omega_C/\partial k_x=0.474\,J_x\) the packet
drifts ballistically by about \(9.5\) cells during one cycle of \(T\approx 0.8\)
ns, on top of which the quantised transport is one cell. Allowing for injection
and collection regions, a chain of at least \(100\) unit cells, that is \(200\)
rings, is required in each of the two arms. Fabrication uniformity across that
many rings is what the disorder study of Sec.~\ref{sec:gap_protection} prices:
the integer survives until the gap closes, and at \(W=0.3\) the count is pinned
across the whole ensemble.

Two device-level remarks complete the picture. The sweep of the coupling
dimerisation \(\delta\) is implemented by replacing each fixed directional coupler
with a Mach--Zehnder interferometer whose arms carry electro-optic phase
shifters, the differential arm phase setting the coupling; the staggered detuning
\(\Delta\) is supplied separately by per-ring shifters. These operate at
modulation bandwidths far exceeding the adiabatic rate required
here~\cite{sacher2012breaking}. The frequency selectivity of the same couplers is
what confines the cycle to the clock band, as anticipated in
Sec.~\ref{sec:sfwm_paw_partition}: for a signal--idler separation of several free
spectral ranges and an arm asymmetry of a few hundred micrometres, the arm phases
seen by the two bands differ by of order a radian, which is sufficient to place
the signal band on a trivial loop while the idler executes the full cycle.

Second, the fringe is accumulated over many heralded pairs per filter setting, so
no per-photon timing resolution finer than the pair coherence is required. The
readout is a phase, not an arrival time, which is precisely why it survives the
narrowband photons that Appendix~\ref{app:packet_displacement} shows to carry
only a fractional displacement. Detector timing enters only through the
coincidence gating that identifies pairs, at the nanosecond scale of the photon
lifetime.

\subsection{Feasibility budget}
\label{sec:feasibility_budget}

Three requirements fix the operating window at once, and the full parameter
budget is collected in Appendix~\ref{app:budget}.

Adiabaticity requires \(T\gtrsim 20/J_x\), with the cycle gap \(2RJ_x\). The
prefactor is set by the criterion that actually governs the winding readout,
namely the first-order non-adiabatic phase error, which scales as \(1/T\) and
extrapolates from a measured \(4\times10^{-3}\) rad at \(T=10^{3}/J_x\)
(Appendix~\ref{app:numerics}) to approximately \(0.2\) rad at \(T=20/J_x\), small
against the \(2.19\) rad unwrapping margin. The transported charge of a uniformly
weighted band at the same cycle time is quantised only at the percent level and
reaches the \(10^{-3}\) level near \(T\approx 50/J_x\); that slower convergence
prices the transport benchmark of Eq.~\eqref{eq:thouless_transport}, not the
fringe, and it is one more respect in which the winding is the cheaper
observable.

One traversal must complete within the photon lifetime
\(\tau_{\mathrm{ph}}=Q/\omega\), as discussed in Sec.~\ref{sec:implementation}.
And the signal filter must step at the scan spacing of Sec.~\ref{sec:sampling}
while holding its passband below the band-edge limit of approximately \(0.15\)
GHz.

Propagation loss is set by the same lifetime and not by the number of rings
traversed. All rings carry the same loaded \(Q\), so the amplitude decays as
\(e^{-t/2\tau_{\mathrm{ph}}}\) with the time spent in the chain, which is \(T\)
by construction, independently of how many cells the photon crosses in that time.
What the lifetime does not cover is coupler excess loss, which is incurred once
per coupler traversed: at roughly ten cells crossed per cycle and an insertion
loss of \(0.01\) dB per Mach--Zehnder coupler, this contributes about \(0.2\) dB
per arm, small against the \(e^{-T/\tau_{\mathrm{ph}}}\) survival.

These requirements do not all relax in the same direction. A larger \(J_x\)
sharpens the gap and shortens the adiabatic cycle, but it also raises the
single-mode correction \((J_x/\mathrm{FSR})^2\) of
Sec.~\ref{sec:rice_mele_clock_band} and shrinks
\(t_{\mathrm{cell}}\propto 1/J_x\). At the conservative operating point, with an
optical frequency of \(193\) THz, a free spectral range of \(18\) GHz,
\(J_x/2\pi=4\) GHz, \(R=0.5\), and \(Q=2.5\times10^{6}\), the adiabatic cycle of
\(T\approx 0.80\) ns sits inside the photon lifetime
\(\tau_{\mathrm{ph}}\approx 2.06\) ns with margin
\(T/\tau_{\mathrm{ph}}\approx 0.39\), the single-mode parameter is
\((J_x/\mathrm{FSR})^2\approx 0.05\), and the tick is
\(t_{\mathrm{cell}}\approx 84\) ps. The group velocity varies around the loop,
from \(0.474\,J_x\) at the reference point to \(0.667\,J_x\) a quarter cycle
later, but the clock family of Eq.~\eqref{eq:clock_time_state} is defined on the
reference Hamiltonian alone, so the tick is fixed there. The budgeted quality
factors are conservative against demonstrated ultra-high-\(Q\) TFLN
microrings~\cite{zhang2017monolithic}.

%
%
%
%
%
%
%

\section{Conclusion and outlook}
\label{sec:conclusion}

A clock degree of freedom defined in the Page--Wootters sense can carry a
topological invariant, in the precise sense fixed in
Sec.~\ref{sec:covariant_clock_states}: the band on which the covariant clock
family is built supports a closed pump cycle whose geometric phase winds by an
integer across the Brillouin zone, and that winding can be measured. The
structural content of one cycle on the clock band is the factorisation of
Eq.~\eqref{eq:factorisation}, a rigid translation of the lattice by \(C\) unit
cells, times a periodic geometric phase, times the dynamical phase. The relation
between the winding and the Chern number is the standard spectral flow of hybrid
Wannier centres; what the factorisation adds is that the same content holds as an
operator identity on the whole band, exact in the adiabatic limit and assuming
nothing about the state on which it acts.

That last property is what fixed the choice of observable. The same integer
appears both as a translation of the lattice and as the winding of
\(\gamma(k_x)\), but the translation acts in full only on a state of uniform
weight across the zone, which a heralded narrowband source does not prepare; the
photon the experiment holds follows the local Berry curvature instead and is
displaced by the fractional value computed in
Appendix~\ref{app:packet_displacement}. Building the measurement on the phase
representation of the same integer is what turns the construction into a
protocol: the winding of the cycle phase across the Brillouin zone, isolated by a
forward and backward pump interferometer that cancels the dynamical phase,
scanned nonlocally through the entangled partner, and counted by unwrapping. The
entanglement is operational at this point rather than interpretational, since an
interferometric phase requires a reference and the partner photon supplies it in
place of an external optical one.

The protection comes in two layers that are logically independent. The ideal
winding is pinned by the gap: it cannot change under any smooth deformation of
the device, of the pump loop, or of the coupler-induced loop family, unless the
gap closes somewhere on the Bloch torus, a statement verified here realisation by
realisation. The readout of that winding is protected by its digital character,
with errors exponentially rare below the threshold set by the sampling margin, in
contrast with the linear degradation of any analog delay estimate. Neither layer
is unconditional, and both are stated as inequalities with their conditions
attached.

Three limits should be kept in view. First, what is protected is the integer and
not the tick duration \(t_{\mathrm{cell}}=a/|v_g|\), which is device dependent
and must be calibrated once, in the way that the SI second fixes a count rather
than a duration; the readout of Sec.~\ref{sec:winding_readout} does not pass
through it. Second, the construction is a proposal: its building blocks are
individually demonstrated in thin-film lithium niobate, but their combination
into a working winding readout has not been realised, and the budget of
Sec.~\ref{sec:feasibility_budget} closes only at the conservative operating
point. Third, the narrowband idealisation underlying the shifted constraint of
Eq.~\eqref{eq:shifted_constraint} incurs corrections of order \(1/Q\), negligible
at the budgeted quality factor but a genuine degradation for low-\(Q\) devices.

Several extensions follow. Promoting the slow control \(\tau\) to a second
internal clock would remove the last appeal to an external time scale, closing
the construction on itself. A higher Chern number would wind the fringe by more
turns per scan, improving the ratio of signal to unwrapping margin. A non-Abelian
generalisation, in which a degenerate clock band is transported by a
matrix-valued holonomy~\cite{sun2022nonabelian}, could extend the construction to
a non-commuting set of protected advances. And the demonstration that quantised
pumping can survive fast driving through tailored
dissipation~\cite{fedorova2020observation} suggests a route to relax the lifetime
constraint of Sec.~\ref{sec:feasibility_budget}, with the caveat that the
engineered loss must not remove or dephase the heralded photon on which both the
conditioning and the readout depend.

One extension bears on the gap between what is quantised and what a heralded
photon can display. The obstacle is that a narrow momentum window samples the
Berry curvature locally, and adding a linear potential removes it by a different
route: under a tilt the momentum is swept across the zone in time, so a packet
that stays narrow at every instant nevertheless samples the full zone over a
Bloch period, and the transport recovers its quantised value for an arbitrary
initial state on the band~\cite{ke2020topological}. Whether that mechanism can be
made to coexist with the covariant clock family, with the adiabatic cycle, and
with the photon lifetime of Sec.~\ref{sec:feasibility_budget} is not settled
here, but it is the concrete route by which the quantised translation, rather
than its phase representation, could become the measured quantity.

\bibliographystyle{quantum}

\newpage

%
%
\onecolumn
\appendix

\section{Page--Wootters partition with the SFWM interaction}
\label{app:sfwm_paw_constraint}

This appendix derives the conditional dynamics of the signal photon upon
conditioning on the idler clock, starting from the full nonlinear Hamiltonian of
the spontaneous four-wave-mixing process. The argument follows the framework of
Smith and Ahmadi~\cite{smith2019quantizing}, specialised to the photonic SFWM
Hamiltonian on the Rice--Mele ring lattice.

\subsection{The full nonlinear Hamiltonian}

In the undepleted-pump regime the pump field at frequency \(\omega_p\) is treated
as a strong classical mode and is not quantised. The remaining quantum degrees of
freedom are the signal field \(\hat a_s\) and the idler field \(\hat a_i\), each
carrying its own copy of the Rice--Mele ring lattice. The total Hamiltonian
decomposes as
\begin{equation}
\label{eq:app_total_hamiltonian}
\hat H_{\mathrm{tot}}=\hat H_C+\hat H_S+\hat H_{\mathrm{NL}} ,
\end{equation}
where \(\hat H_C\) acts on the idler Hilbert space, \(\hat H_S\) on the signal
Hilbert space, and
\begin{equation}
\label{eq:app_nl_coupling}
\hat H_{\mathrm{NL}}
=
g\,\hat a_s^{\dagger}\hat a_i^{\dagger}e^{-i2\omega_p t}+\mathrm{h.c.}
\end{equation}
is the SFWM coupling, with \(g\) set by the third-order susceptibility, the pump
amplitude, and the ring mode overlap. Both \(\hat H_C\) and \(\hat H_S\) are
given by the Bloch Hamiltonian of Eq.~\eqref{eq:rice_mele_hamiltonian} acting on
the respective photon. As discussed in Sec.~\ref{sec:implementation}, the
frequency selectivity of the Mach--Zehnder couplers makes the effective lattice
parameters frequency dependent, so the clock and system photons evolve under
\(\hat H_C=H_{\mathrm{sp}}(k_x;\delta(\omega_i),\Delta(\omega_i))\) and
\(\hat H_S=H_{\mathrm{sp}}(k_x;\delta(\omega_s),\Delta(\omega_s))\); the
derivation below does not require the two parameter sets to be equal.

The explicit time dependence in Eq.~\eqref{eq:app_nl_coupling} is the obstacle to
the naive Page--Wootters construction, which assumes a time-independent global
Hamiltonian.

\subsection{Removing the pump phase}

Transform to the rotating frame defined by
\begin{equation}
\label{eq:app_rotating_frame}
\hat U_p(t)=\exp\!\left[-i\omega_p t\left(\hat N_s+\hat N_i\right)\right] ,
\end{equation}
with \(\hat N_{s,i}\) the photon-number operators on the signal and idler
subspaces. Under this transformation the operators acquire phases
\(\hat a_{s,i}\to e^{-i\omega_p t}\hat a_{s,i}\), so the bilinear
\(\hat a_s^{\dagger}\hat a_i^{\dagger}\) picks up a factor \(e^{+i2\omega_p t}\)
that cancels the explicit time dependence of
Eq.~\eqref{eq:app_nl_coupling} exactly:
\begin{equation}
\label{eq:app_nl_rotating}
\hat H_{\mathrm{NL}}^{(\mathrm{rot})}
=
g\,\hat a_s^{\dagger}\hat a_i^{\dagger}+\mathrm{h.c.}
\end{equation}
The price is a constant shift \(-\omega_p\hat N_{s,i}\) of each free Hamiltonian.
Since the relevant Fock subspace contains exactly one signal and one idler photon,
this evaluates to \(-2\omega_p\) in total, and
\begin{equation}
\label{eq:app_total_rotating}
\hat H_{\mathrm{tot}}^{(\mathrm{rot})}
=
\hat H_C+\hat H_S+\hat H_{\mathrm{NL}}^{(\mathrm{rot})}-2\omega_p I ,
\end{equation}
where the identity acts on the single-pair subspace.

\subsection{The shifted constraint}

The Page--Wootters construction requires the physical state to satisfy
\(\hat J|\Psi\rangle\rangle=0\). For the SFWM-coupled system the shifted
constraint of Eq.~\eqref{eq:shifted_constraint} is adopted, which differs from
Eq.~\eqref{eq:app_total_rotating} only by the omission of
\(\hat H_{\mathrm{NL}}^{(\mathrm{rot})}\). The key observation is that the
entangled pair state generated by SFWM, in the undepleted-pump regime, is
annihilated by this constraint to leading order in \(g\). Acting with \(\hat J\)
on the state of Eq.~\eqref{eq:sfwm_state},
\begin{equation}
\label{eq:app_constraint_action}
\begin{aligned}
\hat J|\Psi\rangle\rangle
={}&
\sum_{k_x^{(i)},k_x^{(s)}}
\Big[\omega_C\!\left(k_x^{(i)}\right)+\omega_S\!\left(k_x^{(s)}\right)-2\omega_p\Big]
\Phi
\\
&\times
|u_C(k_x^{(i)})\rangle\otimes|u_S(k_x^{(s)})\rangle
\;\simeq\;0 ,
\end{aligned}
\end{equation}
the bracketed factor vanishing on the phase-matching manifold of
Eq.~\eqref{eq:energy_anticorrelation}. Corrections are of order
\(\Delta\omega/\omega_p\), with \(\Delta\omega\) the joint-spectral bandwidth. For
a cavity-limited microring source this bandwidth is set by the loaded ring
linewidth, so the fractional constraint violation is of order \(1/Q\), far below
any scale entering the winding.

The nonlinear term is therefore responsible for generating the entangled state
rather than appearing in the constraint that the state satisfies. This is the
physical meaning of the shift: the SFWM coupling shapes the joint spectrum, after
which the resulting state evolves under the non-interacting constraint.

\subsection{Conditional dynamics on the signal}

Define the conditional signal state as in
Eq.~\eqref{eq:conditional_signal_state}. With the clock family of
Eq.~\eqref{eq:clock_time_state}, whose weight \(w(k_x)\) is independent of \(t\),
the dual bra satisfies
\begin{equation}
\label{eq:app_dual_bra}
{}_C\langle t|
=
\int_{\mathcal B_C}\! dk_x\,w(k_x)\,e^{+i\omega_C(k_x)t}\,\langle u_C(k_x)| ,
\qquad
\partial_t\,{}_C\langle t|=+i\,{}_C\langle t|\hat H_C ,
\end{equation}
the second relation holding for any choice of weight. Using the constraint to
replace \(\hat H_C\) acting on the clock factor by
\(2\omega_p I-\hat H_S\) acting on the system factor, and differentiating
Eq.~\eqref{eq:conditional_signal_state},
\begin{equation}
\label{eq:app_conditional_evolution}
i\frac{d}{dt}|\psi_S(t)\rangle
=
-\,{}_C\langle t|\hat H_C|\Psi\rangle\rangle
=
\left(\hat H_S-2\omega_p\right)|\psi_S(t)\rangle .
\end{equation}
The constant contributes only an overall phase, which is factored out by writing
\(|\psi_S(t)\rangle=e^{+i2\omega_p t}|\tilde\psi_S(t)\rangle\), giving
\begin{equation}
\label{eq:app_schrodinger}
i\frac{d}{dt}|\tilde\psi_S(t)\rangle=\hat H_S|\tilde\psi_S(t)\rangle ,
\end{equation}
which is exactly Schr\"odinger evolution under \(\hat H_S\). No reflection
\(t\to-t\) or alternative sign convention is required.

\subsection{The global phase}

The factor \(e^{+i2\omega_p t}\) separated above is a global phase on the signal
state. It cancels in all observables on the signal alone, including local
intensities and homodyne measurements. It does appear in any joint measurement
that compares signal and idler phases, but in the interferometer of
Sec.~\ref{sec:winding_readout} it is common to the forward and backward
traversals, which differ only in loop orientation and not in pump frequency, so
it cancels identically in the fringe difference of
Eq.~\eqref{eq:fringe_difference} along with all other common-mode phases. It does
not enter the winding.

\section{Gauge structure, winding, and the factorisation}
\label{app:gauge_winding}

\subsection{Adiabatic phase at fixed momentum}

At fixed \(k_x\) the adiabatic theorem for the isolated lower band gives, over
one closed cycle,
\begin{equation}
\label{eq:app_adiabatic_phase}
\hat U_{\mathrm{cycle}}|u(k_x,0)\rangle
=
e^{i\gamma(k_x)}
e^{-i\int_0^T\omega_C(k_x,\tau)d\tau}
|u(k_x,0)\rangle ,
\end{equation}
with \(\gamma(k_x)\) the Berry phase of Eq.~\eqref{eq:berry_phase}, defined modulo
\(2\pi\) and gauge invariant because the \(\tau\) loop is closed. Non-adiabatic
corrections enter the phase at order \(1/T\) and the amplitude at order
\(1/T^2\); the \(1/T\) phase scaling is verified numerically in
Appendix~\ref{app:numerics}.

\subsection{Winding of the geometric phase}

Cut the \((k_x,\tau)\) torus along \(k_x\) and choose, on the resulting cylinder,
a gauge for \(|u(k_x,\tau)\rangle\) that is periodic in \(\tau\) and smooth in
both arguments; such a gauge always exists on the cylinder, only the torus being
able to obstruct it. Then
\begin{equation}
\label{eq:app_stokes}
\begin{aligned}
2\pi C
={}&
-\int_{-\pi}^{\pi}\!\! dk_x\!\int_0^T\!\! d\tau
\left(\partial_{k_x}A_\tau-\partial_\tau A_{k_x}\right)
\\
={}&
-\left[\gamma(\pi)-\gamma(-\pi)\right]
\\
&+\int_{-\pi}^{\pi}\!\!\left[A_{k_x}(k_x,T)-A_{k_x}(k_x,0)\right] dk_x ,
\end{aligned}
\end{equation}
where the overall sign matches the transport orientation of
Eq.~\eqref{eq:chern_number}. The second term vanishes by the \(\tau\) periodicity
of the gauge, leaving Eq.~\eqref{eq:winding_relation}.

When \(C\neq 0\) the smooth cylinder gauge cannot in addition be periodic in
\(k_x\), which is the standard statement that no global smooth gauge exists on a
torus of nonzero Chern number. The winding of \(\gamma\) is gauge invariant
regardless, because two admissible gauges differ in \(\gamma(k_x)\) by a smooth
single-valued function of \(k_x\) plus multiples of \(2\pi\), and neither can
change an unwrapped closed-loop winding.

\subsection{Proof of the factorisation}

Split \(\gamma(k_x)=-Ck_x+\gamma_{\mathrm{per}}(k_x)\), where
Eq.~\eqref{eq:winding_relation} guarantees
\(\gamma_{\mathrm{per}}(\pi)=\gamma_{\mathrm{per}}(-\pi)\) so that
\(\gamma_{\mathrm{per}}\) is a legitimate periodic function of \(k_x\). Inserting
into Eq.~\eqref{eq:app_adiabatic_phase}, the operator on the band reads
\begin{equation}
\label{eq:app_factorisation}
\hat U_{\mathrm{cycle}}
=
e^{-iC\hat k_x}\,
e^{i\gamma_{\mathrm{per}}(\hat k_x)}\,
e^{-i\int_0^T\omega_C(\hat k_x,\tau)d\tau} ,
\end{equation}
and \(e^{-iC\hat k_x}\) is the translation \(\hat T_{Ca}\), since it maps
\(\psi(x)\) to \(\psi(x-Ca)\), moving a packet forward by \(Ca\). This is
Eq.~\eqref{eq:factorisation}. Every factor is diagonal in \(k_x\), so the
decomposition is simultaneously an operator identity and a statement about each
momentum fibre.

%
%
%
%
%

\section{Displacement of a narrow packet}
\label{app:packet_displacement}

Let the initial state be a lower-band packet
\(|\Psi(0)\rangle=\int dk_x\,f(k_x)|u(k_x,0)\rangle\) with \(|f|^2\) normalised
and concentrated near \(k_0\). By Eq.~\eqref{eq:app_adiabatic_phase},
\begin{equation}
\label{eq:app_evolved_packet}
|\Psi(T)\rangle
=
\int dk_x\, f(k_x)\,e^{i\Phi_{\mathrm{tot}}(k_x)}|u(k_x,0)\rangle ,
\end{equation}
with
\(\Phi_{\mathrm{tot}}=\gamma(k_x)-\int_0^T\omega_C(k_x,\tau)\,d\tau\). Since
\(f\) and the Berry connection of the band basis are the same before and after
the cycle, the displacement of the centre of mass is carried entirely by the
accumulated phase:
\begin{equation}
\label{eq:app_displacement_law}
\Delta x
=
-\int dk_x\,|f(k_x)|^2\,\partial_{k_x}\Phi_{\mathrm{tot}}(k_x)
=
\int dk_x\,|f|^2
\left[-\partial_{k_x}\gamma+\int_0^T\!\frac{\partial\omega_C}{\partial k_x}d\tau\right] .
\end{equation}
For a packet narrow against the scale of variation of the integrands the
geometric term reduces to the local slope of \(\gamma\) at the occupied momentum.
At the operating point \(k_x=\pi/2\) the computed value is
\begin{equation}
\label{eq:local_slope}
-\left.\partial_{k_x}\gamma\right|_{\pi/2}=+0.318\ \text{cells},
\end{equation}
so a heralded packet is displaced by a fraction of a cell rather than by \(C\).

\begin{figure*}[t]
\centering
\includegraphics[width=\textwidth]{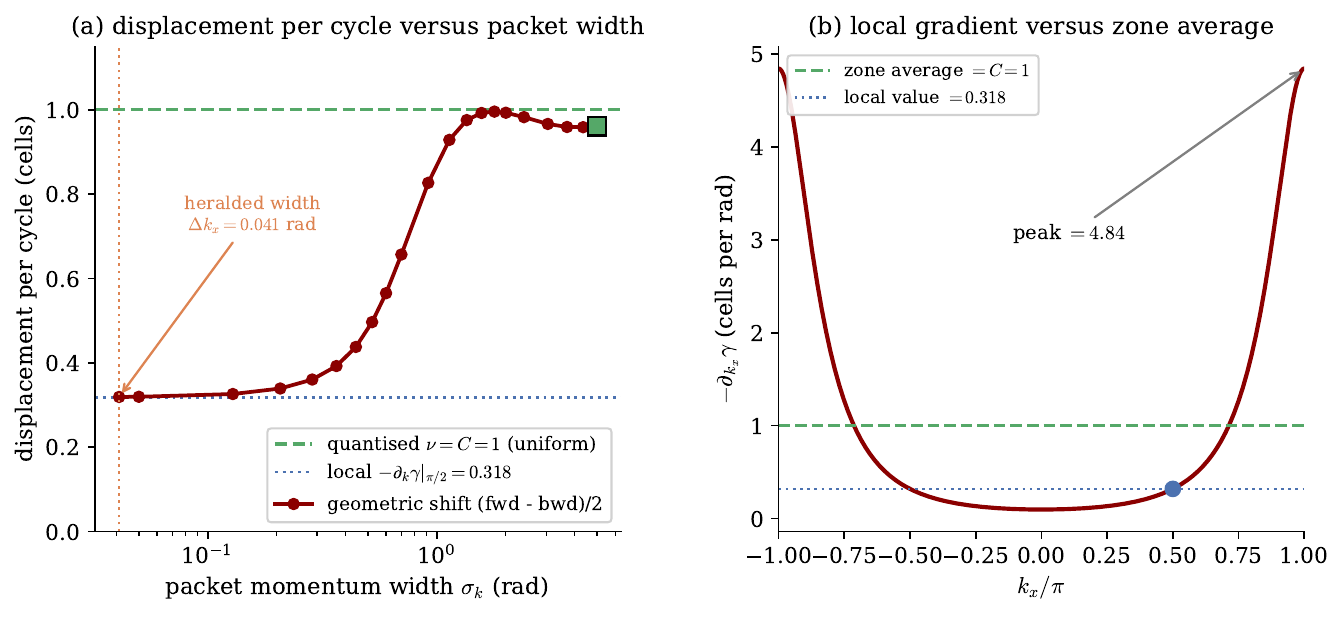}
\caption{%
(a) Displacement per cycle of a lower-band packet against its momentum width,
with the geometric part isolated as half the difference between forward and
backward traversals.  At the heralded width \(\Delta k_x=0.041\) rad the shift
is \(0.319\) cells; it rises continuously as the occupied window widens, is
fractional at every finite width, and reaches \(C\) only at uniform weight
across the zone.  The approach is not monotone near the wide end, the weighted
average of the phase gradient overshooting slightly before the uniform limit.
(b) Gradient of the geometric phase across the zone, against its zone average.
The local value at the operating point is a factor of three below the average
of \(C=1\), and the gradient concentrates near \(k_x=\pi\) where the pump loop
passes closest to the degeneracy.  Only the zone average of this curve is
quantised, which is why the readout scans the phase rather than measuring a
displacement.
}
\label{fig:packet}
\end{figure*}

Equation~\eqref{eq:app_displacement_law} also establishes the two limits quoted
in the main text. The dynamical term is the zone average of the group velocity,
and the integral of the derivative of a periodic function over the zone vanishes
identically,
\begin{equation}
\label{eq:zone_average_vg}
\frac{1}{2\pi}\int_{-\pi}^{\pi}\frac{\partial\omega_C}{\partial k_x}\,dk_x=0 ,
\end{equation}
so for uniform weight only the geometric term survives, and it is
\(-\frac{a}{2\pi}\oint\partial_{k_x}\gamma\,dk_x=+Ca\) by
Eq.~\eqref{eq:winding_relation}: a state of uniform weight advances by exactly
\(C\) cells. Such a state does not propagate, however, carrying equal
right-moving and left-moving weight, which is the \(\pm k_x\) degeneracy of
Sec.~\ref{sec:covariant_clock_states} seen from the transport side. For any
non-uniform weight the geometric term is a weighted average of a smooth
non-constant function, and has no reason to be, and numerically is not, an
integer.

Two remarks connect this to the rest of the paper. First, under path reversal
\(\gamma\to-\gamma\) while the dynamical integrand is unchanged, so half the
difference of the forward and backward displacements isolates the geometric part.
This is the estimator used in the packet numerics of
Appendix~\ref{app:numerics}, and it is the real-space shadow of the
interferometric fringe of Eq.~\eqref{eq:fringe_difference}. Second, the covariant
clock family of Eq.~\eqref{eq:clock_time_state} carries the density-of-states
weight of Eq.~\eqref{eq:povm_weight} and is therefore not the uniformly weighted
object either. The quantised observable of this paper is the winding of
Eq.~\eqref{eq:winding_relation}, which is a property of the cycle and not of any
prepared state, and the readout of Sec.~\ref{sec:winding_readout} never requires
a state of uniform weight to be prepared.

\section{Fringe algebra of the two-traversal interferometer}
\label{app:fringe_algebra}

\subsection{The ideal fringe}

Consider one heralded pair, with the signal filtered at the setting that heralds
clock momentum \(k_x\), and let the pump control be driven through the forward
loop in one arm of the interferometer and through the backward loop in the other,
with a controllable relative phase \(\theta_{\mathrm{ref}}\) between the arms.
Using Eq.~\eqref{eq:app_adiabatic_phase} and its backward counterpart, in which
the dynamical phase is invariant and \(\gamma\to-\gamma\), the pair amplitude
reaching the coincidence detectors is
\begin{equation}
\label{eq:app_pair_amplitude}
A(k_x)
\propto
e^{i\left[\phi_{\mathrm{dyn}}+\gamma(k_x)\right]}e^{i\chi}
+
e^{i\left[\phi_{\mathrm{dyn}}-\gamma(k_x)\right]}e^{i\left(\chi+\theta_{\mathrm{ref}}\right)} ,
\end{equation}
where \(\chi\) collects every phase common to the two arms: the pump laser phase,
path lengths shared by the arms, and the global \(2\omega_p t\) phase of
Appendix~\ref{app:sfwm_paw_constraint}. The coincidence rate is
\begin{equation}
\label{eq:app_coincidence_rate}
R_{\mathrm{cc}}(k_x)
\propto
1+\cos\!\left[2\gamma(k_x)-\theta_{\mathrm{ref}}\right] ,
\end{equation}
so sweeping \(\theta_{\mathrm{ref}}\) at each filter setting traces a fringe whose
phase is \(2\gamma(k_x)\), with \(\phi_{\mathrm{dyn}}\) and \(\chi\) cancelled
identically. This is Eq.~\eqref{eq:fringe_difference} in operational form.

The signal photon plays two roles at once: its filtered frequency selects \(k_x\)
through the anticorrelation of Eq.~\eqref{eq:energy_anticorrelation}, and its
amplitude is the reference against which the idler's differential phase is read.
Noise in \(\chi\), however large, drops out of
Eq.~\eqref{eq:app_coincidence_rate}. What remains as the fringe phase noise
\(\sigma_\varphi\) of Sec.~\ref{sec:digital_protection} is the shot noise of the
finite coincidence count at each setting together with any arm-differential drift
over the acquisition, both of which enter \(2\gamma\) directly and are the
quantities that the threshold inequality of Eq.~\eqref{eq:threshold_bound}
prices.

\subsection{Incomplete separation of the two loops}
\label{app:cc_cs}

Section~\ref{sec:sfwm_paw_partition} designs the device so that the idler loop
encircles the degeneracy while the signal loop does not. Suppose the separation
is incomplete and the signal band acquires a Chern number \(C_S\neq0\). One
modulation waveform drives one physical chain, so both photons traverse their
respective loops in the same sense, and reversing the waveform reverses both
orientations together. The two geometric phases therefore enter
Eq.~\eqref{eq:app_pair_amplitude} with the same sign,
\begin{equation}
\label{eq:app_two_band_phases}
\begin{aligned}
\Phi_{\mathrm{fwd}}&=\phi_{\mathrm{dyn}}+\gamma_C(k_i)+\gamma_S(k_s),
\\[4pt]
\Phi_{\mathrm{bwd}}&=\phi_{\mathrm{dyn}}-\gamma_C(k_i)-\gamma_S(k_s),
\end{aligned}
\end{equation}
and the fringe carries their sum,
\(\Delta\Phi=2\left[\gamma_C(k_i)+\gamma_S(k_s)\right]\).

What converts that sum into a difference of windings is the anticorrelation
itself. By Eq.~\eqref{eq:dispersion_closed_form} the lower band is monotonically
increasing in \(k\) on the directional branch, for the signal as well as for the
idler. The constraint \(\omega_C(k_i)+\omega_S(k_s)=2\omega_p\) then forces
\(dk_s/dk_i<0\): stepping the filter so that \(k_i\) advances across its zone
sweeps \(k_s\) backward across its own. Over one complete scan the two windings
accumulate with opposite signs,
\begin{equation}
\label{eq:app_winding_accumulation}
\left[\gamma_C\right]_{k_i=-\pi}^{k_i=+\pi}=-2\pi C_C ,
\qquad
\left[\gamma_S\right]_{\text{same scan}}=+2\pi C_S ,
\end{equation}
the second differing in sign from Eq.~\eqref{eq:winding_relation} because the
traversal of \(k_s\) is orientation reversing. The measured winding is therefore
\begin{equation}
\label{eq:app_measured_winding}
W=\frac{1}{2\pi}\left[\Delta\Phi\right]_{k_i=-\pi}^{k_i=+\pi}
=-2\left(C_C-C_S\right) ,
\end{equation}
which reduces to Eq.~\eqref{eq:winding_readout} when \(C_S=0\). The estimator
returns an even integer in either case, so the parity flag of
Sec.~\ref{sec:momentum_sign} is unaffected.

Two qualifications belong with this result. The sum-to-difference conversion
rests on both bands being monotonically increasing on their populated branches;
an inverted signal band would restore the sum, and the design must fix which
branch is populated on each side. And an integer answer is not the same as the
intended one: \(C_C-C_S=1\) is returned by \((1,0)\) and by \((2,1)\) alike, so
Eq.~\eqref{eq:app_measured_winding} shows that imperfect separation cannot
produce a fractional result, not that it can be left uncontrolled. The loop
separation of Sec.~\ref{sec:sfwm_paw_partition} remains a design requirement.

\subsection{Two physically distinct chains}
\label{app:two_chains}

In the parallel implementation of Sec.~\ref{sec:implementation} the two arms are
separate chains, so the signal propagation phase is no longer common to them by
construction, and the term \(\chi\) of Eq.~\eqref{eq:app_pair_amplitude} splits
into \(\chi_A\) and \(\chi_B\). The residual
\(\delta\chi(k_s)=\chi_A(k_s)-\chi_B(k_s)\) is set by the fabrication difference
between the two chains and does not cancel.

It does not enter the winding. Both chains are gapped over the whole scan, so
each propagation phase is a smooth single-valued function of \(k_s\), periodic
across the zone; their difference is therefore also smooth, single-valued and
periodic, and a periodic function accumulates exactly zero winding around the
closed scan. This is the same argument that disposes of the residual dynamical
phase \(\delta\phi_{\mathrm{dyn}}\) in Sec.~\ref{sec:implementation}, applied to
the reference arm rather than to the clock arm.

What \(\delta\chi\) does cost is unwrapping margin, through its gradient. For
percent-level coupling mismatch between the chains the propagation phase differs
by of order \(\delta J_x T\approx0.2\) rad per radian of \(k_s\), and the
Jacobian \(|dk_s/dk_i|\) of the anticorrelation is of order unity when the two
bands have comparable widths, so the contribution to the fringe step is
comparable to that of \(\delta\phi_{\mathrm{dyn}}\) and small against the
\(2.19\) rad margin of Sec.~\ref{sec:sampling}. A constant offset between the
chains shifts every sample equally and drops out of the differences the estimator
of Eq.~\eqref{eq:winding_estimator} forms.

%
%
%
%
%
%

\section{Numerical methods and results}
\label{app:numerics}

All computations use the closed two-level form of the propagator,
\begin{equation}
\label{eq:app_propagator}
e^{-i\mathbf d\cdot\boldsymbol\sigma\,dt}
=
\cos(|\mathbf d|dt)\,I
-i\sin(|\mathbf d|dt)\,\frac{\mathbf d\cdot\boldsymbol\sigma}{|\mathbf d|} ,
\end{equation}
which is exact per step.

\subsection{The invariant and the geometric-phase profile}

The invariant of the cycle is computed three independent ways: the lattice
field-strength method on the \((k_x,\tau)\) torus, the direct integral of
Eq.~\eqref{eq:chern_number}, and the net advance of the Wannier centre over one
cycle. All give \(|C|=1\), with the transport in \(+x\) fixing \(C=+1\) in the
orientation of Sec.~\ref{sec:cycle_chern}.

\begin{figure*}[t]
\centering
\includegraphics[width=\textwidth]{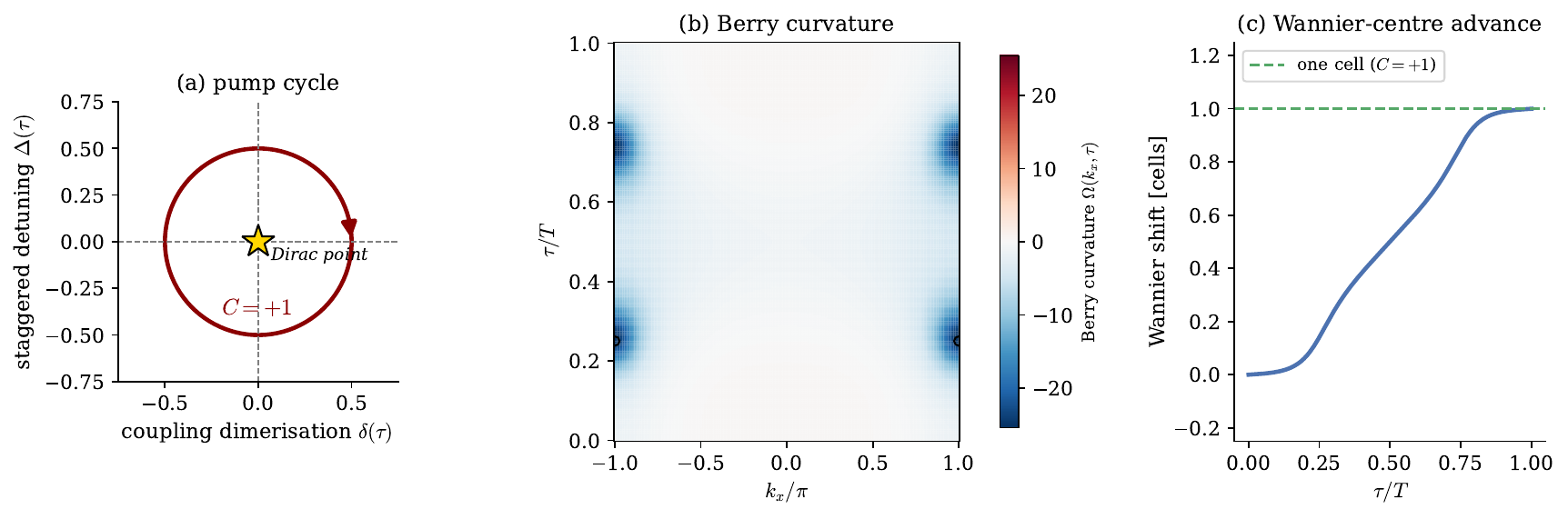}
\caption{%
(a) The loop of Eq.~\eqref{eq:pump_cycle} in the \((\delta,\Delta)\) plane.
(b) Berry curvature of the lower band over the \((k_x,\tau)\) torus. It
concentrates near the zone edges at \(\tau/T=1/4\) and \(3/4\), where the loop
passes closest to the degeneracy; this is also why
\(\partial_{k_x}\gamma\) peaks at \(k_x=\pi\).
(c) Wannier centre through one cycle, advancing by one unit cell.
}
\label{fig:berry}
\end{figure*}

The profile \(\gamma(k_x)\) is computed as a Wilson loop over the closed \(\tau\)
cycle at each \(k_x\): the phase of the product of normalised overlaps between
lower-band eigenstates at successive \(\tau\) grid points. This estimator is
gauge invariant and exact for eigenstates, with no adiabaticity requirement. It
is also independent of where on the cycle the dispersion is evaluated, since it
integrates over the whole loop. On a \(720\times1200\) grid the winding across
the zone is \(-1.000000\) turns, the parity relation of
Eq.~\eqref{eq:gamma_parity} holds to a maximum residual of
\(3.6\times10^{-15}\) rad, the half-zone winding including the \(k_x=\pi\)
endpoint is \(-0.500000\) turns, and the peak gradient is \(4.844\) at
\(k_x=\pi\), with the value \(0.318\) at the operating point \(\pi/2\).

A real-time cross-check evolves the lower eigenstate through one cycle at
\(T=10^{3}/J_x\) and subtracts the accumulated dynamical phase. The residual
difference from the Wilson-loop value is at the \(10^{-4}\) to
\(4\times10^{-3}\) rad level across the zone and scales as \(1/T\), verified by
halving: the error ratio between \(T=300/J_x\) and \(T=10^{3}/J_x\) is \(3.33\).
This identifies it as the first-order non-adiabatic phase correction rather than a
numerical artefact. The forward and backward structure used by the protocol is
verified in the same runs: the dynamical phase is identical in the two traversal
directions to \(10^{-14}\), and the fringe difference equals \(2\gamma(k_x)\) to
the same \(1/T\) accuracy.

\subsection{Filled-band transport}

\begin{figure}[t]
\centering
\includegraphics[width=\columnwidth]{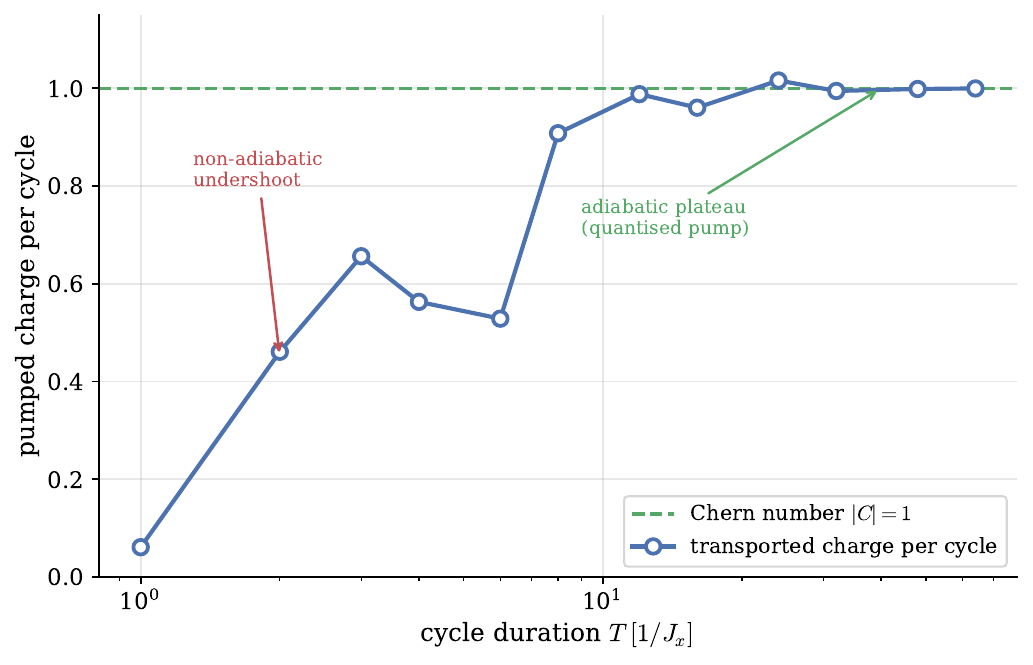}
\caption{%
Transported charge per cycle against cycle duration. The approach to the
quantised value is not monotone, Landau--Zener interference being superposed on
the convergence, and the \(10^{-3}\) level is reached only near
\(T\approx 50/J_x\). The fringe phase converges faster, which is why the
adiabaticity prefactor in the budget is set by the phase error rather than by
this benchmark.
}
\label{fig:adiabatic}
\end{figure}

The filled-band transport is computed by filling the lower band on a finite
momentum ring and evolving each Bloch state under
\(H_{\mathrm{sp}}(k_x;\delta(\tau),\Delta(\tau))\) through one cycle of duration
\(T\), the transported charge being extracted as the change in the
King-Smith--Vanderbilt polarisation~\cite{kingsmith1993theory}. For fast cycles
the charge falls short and oscillates, with Landau--Zener-type interference
superposed on the convergence: the deviation from \(|C|=1\) is
\(1.6\times10^{-2}\) at \(T=24/J_x\), \(5\times10^{-3}\) at \(32/J_x\),
\(1.4\times10^{-3}\) at \(48/J_x\), and \(2\times10^{-4}\) at \(64/J_x\). The
\(10^{-3}\) level is therefore reached only near \(T\approx 50/J_x\), and the
approach is not monotone, the charge overshooting slightly to \(1.016\) at
\(T=24/J_x\) before settling. As noted in Sec.~\ref{sec:feasibility_budget}, this
benchmark is slower to converge than the fringe phase and does not set the
adiabaticity prefactor used in the budget.

\subsection{Packet displacement}

Two independent computations produce the packet results of
Appendix~\ref{app:packet_displacement}. The first is direct: a Gaussian
lower-band packet of momentum width \(\sigma_k\) centred at \(k_x=\pi/2\) is
built on a ring of \(240\) cells and evolved through one full cycle at
\(T=200/J_x\). The clean lattice conserves \(k_x\), so each Bloch spinor evolves
under Eq.~\eqref{eq:app_propagator} and real space enters only in reading off the
centre of mass, with sublattice offsets \(x_A=m\), \(x_B=m+\tfrac12\) and
circular statistics for the ring coordinate. The geometric part of the
displacement is isolated as half the difference between forward and backward
traversals. The second uses the stationary-phase law of
Eq.~\eqref{eq:app_displacement_law}, with \(\gamma\) taken from the Wilson-loop
profile.

At \(\sigma_k=0.1\) the two agree to \(2\times10^{-4}\) cells, \(0.3229\) against
\(0.3231\), validating the law. The width sweep then uses the exact
weighted-gradient form, giving \(0.319\) at the heralded width
\(\sigma_k=0.041\), rising through \(0.37\), \(0.57\), and \(0.87\) at
\(\sigma_k=0.3\), \(0.6\), and \(1.0\), overshooting slightly to \(1.010\) and
\(1.021\) at \(\sigma_k=1.6\) and \(2.5\), and equalling \(1.000000\) at exactly
uniform weight. The sweep is flat at small \(\sigma_k\), converging to the local
slope \(0.318\) of Eq.~\eqref{eq:local_slope}, so the displacement at the
heralded width is insensitive to the precise value of \(\Delta k_x\). The
zone-averaged group velocity vanishes to \(6\times10^{-16}\), confirming
Eq.~\eqref{eq:zone_average_vg}.

\subsection{Deformation ensemble and disorder}

The ensemble of Eq.~\eqref{eq:deformation_ensemble} uses, for each Cartesian
component of \(\mathbf f\), an independent random truncated Fourier series on the
torus with modes up to order three in each direction and unit root-mean-square
amplitude, so that the deformation is smooth, periodic, and generically breaks
every symmetry of the clean cycle, including the parity of
Eq.~\eqref{eq:gamma_parity}. For each of \(40\) realisations and \(13\) strengths
\(W\in[0,0.6]\), the winding of the deformed \(\gamma(k_x)\) is computed by
Wilson loops on a \(360\times480\) torus grid, and the minimum of
\(|\mathbf d|\) over the torus is tracked along the deformation ray at \(80\)
intermediate strengths. Classifying realisations by whether that path minimum
ever fell below \(0.03\,J_x\), the grid resolution scale, yields the dichotomy
reported in Sec.~\ref{sec:gap_protection}: all \(409\) path-gapped
realisation-strength pairs have winding \(-1\) to \(3\times10^{-16}\), and every
changed winding is paired with a path-gap collapse.

\begin{figure}[t]
\centering
\includegraphics[width=\columnwidth]{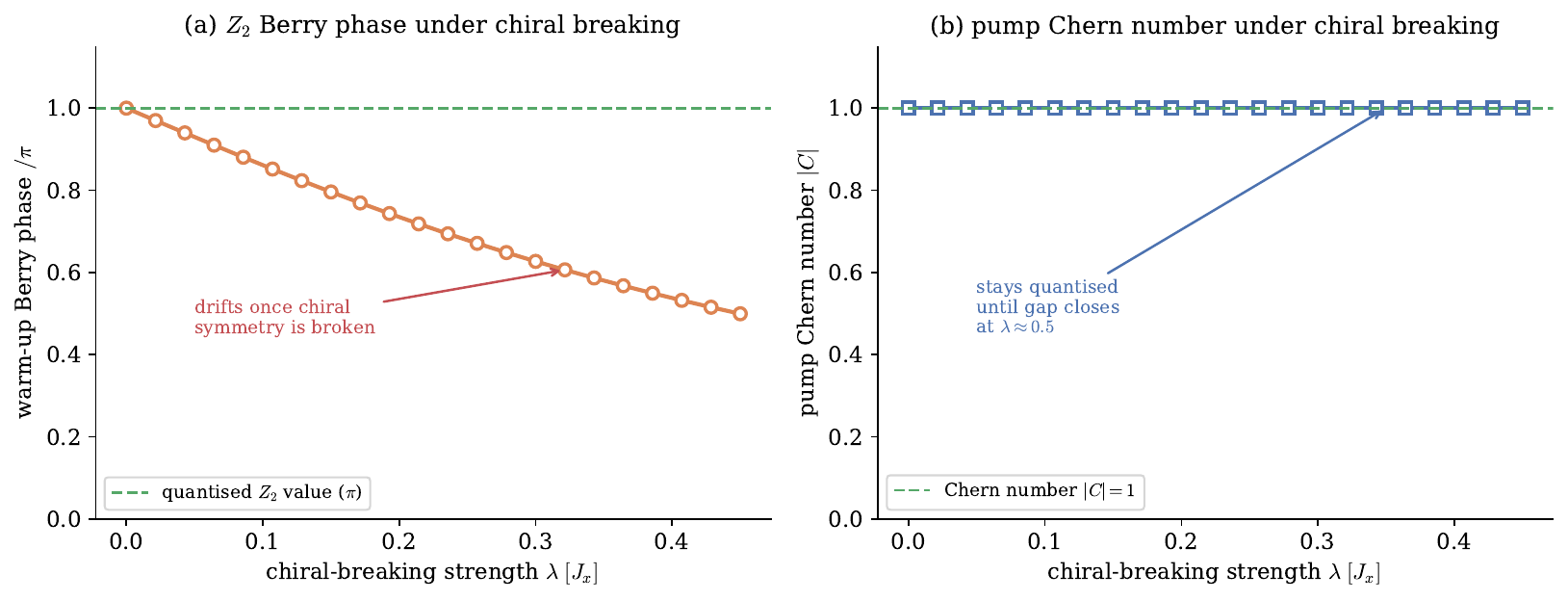}
\caption{%
Protection by symmetry against protection by a gap, under a
chiral-breaking term \(\lambda\,\sigma_z\) held fixed outside the cycle.
(a) A \(Z_2\) Berry phase, quantised by chiral symmetry, drifts continuously as
soon as that symmetry is broken. (b) The Chern number of the pump is unmoved
until \(\lambda\) closes the gap near \(0.5\,J_x\), beyond which it is no longer
defined.
}
\label{fig:symmetry}
\end{figure}

The real-space counterpart uses quenched multiplicative coupling disorder and
additive detuning disorder of strength \(W\), with the pumped count computed for
each realisation by Resta polarisation transport~\cite{resta1998quantum} on a
ring, which is exact under periodic boundary conditions. At \(W=0.3\) the count is
pinned at exactly one cell across the ensemble, with a standard deviation at the
level of machine precision, while a non-topological group-velocity delay
reference drifts with a spread set linearly by the disorder. The count remains
exactly quantised as \(W\) grows until the disorder closes the bulk gap. Over the
same ensemble the tick duration \(t_{\mathrm{cell}}\) inherits the disorder of the
group velocity and spreads by several percent.

\subsection{Windowed readout and Monte Carlo}

The finite-window readout of Eq.~\eqref{eq:windowed_phasor} is verified on the
same geometric-phase profile. The passband lineshape is mapped through the
lower-band dispersion at the reference point \((\delta,\Delta)=(R,0)\), where the
clock states of Eq.~\eqref{eq:clock_time_state} are defined and where
\(\partial\omega_C/\partial k_x=0.474\,J_x\) at the band centre; it is then
truncated to the populated half zone, and the estimator of
Eq.~\eqref{eq:winding_estimator} is applied to the resulting phases on the
\(N_{\mathrm{scan}}=64\) grid, with the two half-zone records stitched through
the parity relation of Eq.~\eqref{eq:gamma_parity}.

At this dispersion the lower band spans \(J_x\), or \(4\) GHz, across the half
zone, so the \(32\) settings per half zone sit at a mean spacing of \(125\) MHz
and the windows overlap at passbands approaching the \(0.15\) GHz ceiling. The
overlap correlates adjacent samples but does not move the winding, which is the
expected behaviour: smoothing cannot change a winding without dragging the phasor
through zero.

The results are collected in Table~\ref{tab:windowed}. The winding is \(-2C\) at
the ring linewidth and at a \(0.15\) GHz passband for Gaussian and Lorentzian
lineshapes alike, with worst-sample visibilities of \(0.83\) and \(0.74\) at the
linewidth. At a \(0.30\) GHz nominal width the two lineshapes separate: the
Gaussian still returns the integer, while the Lorentzian misreads the junction
step at \(k_x=\pi\) and fails by exactly one turn, returning \(-1\), an odd
result that the parity flag of Sec.~\ref{sec:momentum_sign} detects. This is the
numerical content of the apodisation recommendation in Sec.~\ref{sec:sampling}:
Lorentzian tails smear more phase into the window at equal nominal width. The
failure is insensitive to the scan density, confirming it as a window-truncation
bias rather than undersampling.

\begin{table}[t]
\centering
\caption{Windowed readout of Eq.~\eqref{eq:windowed_phasor} at
\(N_{\mathrm{scan}}=64\), with the passband mapped through the dispersion at
\((\delta,\Delta)=(R,0)\).  The ring linewidth at \(Q=2.5\times10^{6}\) is
\(0.077\) GHz.  The raw junction step is the largest wrapped difference before
the re-referencing of Sec.~\ref{sec:momentum_sign}.}
\label{tab:windowed}
\begin{tabular}{lccc}
Passband and lineshape & Winding & Worst \(V_j\) & Raw junction step \\
\hline
Ring linewidth, Gaussian    & \(-2.000\) & 0.83 & 1.68 rad\\
Ring linewidth, Lorentzian  & \(-2.000\) & 0.74 & 1.90 rad\\
0.15 GHz, Gaussian          & \(-2.000\) & 0.72 & 2.25 rad\\
0.15 GHz, Lorentzian        & \(-2.000\) & 0.61 & 2.46 rad\\
0.30 GHz, Gaussian          & \(-2.000\) & 0.58 & 3.05 rad\\
0.30 GHz, Lorentzian        & \(-1.000\) & 0.45 & 3.07 rad\\
\end{tabular}
\end{table}

The Monte Carlo of Sec.~\ref{sec:digital_protection} applies the estimator of
Eq.~\eqref{eq:winding_estimator} to the computed fringe profile with independent
Gaussian phase noise per sample, using \(6000\) trials per noise level.

%
%
%

\section{Feasibility budget}
\label{app:budget}

Tables~\ref{tab:budget_operating} and~\ref{tab:budget_readout} collect the
parameter budget discussed in Sec.~\ref{sec:feasibility_budget}: the first fixes
the operating point and the time scales it implies, the second the device scale
and the requirements the readout places on it.

\begin{table}[t]
\centering
\caption{Operating point and time scales, anchored to demonstrated TFLN ring
platforms.  Coupling rates are angular frequencies; the table quotes
\(J_x/2\pi\) in Hz, while \(T\gtrsim20/J_x\) and
\(\tau_{\mathrm{ph}}=Q/\omega\) use the angular \(J_x\) and \(\omega=2\pi f\).
The group velocity is evaluated at the reference Hamiltonian \(\tau=0\), which by
Eq.~\eqref{eq:pump_cycle} is \((\delta,\Delta)=(R,0)\); there
\(\partial\omega_C/\partial k_x=0.474\,J_x\) at the band centre, so
\(t_{\mathrm{cell}}=1/(0.474\,J_x)\).}
\label{tab:budget_operating}
\begin{tabular}{lc}
Quantity & Value \\
\hline
\multicolumn{2}{l}{\textit{Anchored}}\\
Optical frequency \(f\) & 193 THz\\
Free spectral range & 18 GHz\\
Quality factor \(Q\) & \(2.5\times10^{6}\)\\
EO modulation bandwidth & up to 67 GHz\\
Half-wave voltage \(V_\pi\) & 1.75 V\\
Coupler insertion loss & 0.01 dB\\
\hline
\multicolumn{2}{l}{\textit{Operating point}}\\
Ring-to-ring coupling \(J_x/2\pi\) & 4 GHz\\
Dimerisation radius \(R\) & 0.5\\
Cycle gap \(2RJ_x/2\pi\) & 4 GHz\\
Single-mode parameter \((J_x/\mathrm{FSR})^2\) & 0.05\\
\hline
\multicolumn{2}{l}{\textit{Time scales}}\\
Photon lifetime \(\tau_{\mathrm{ph}}=Q/\omega\) & 2.06 ns\\
Adiabatic cycle time \(T\) & 0.80 ns\\
Lifetime ratio \(T/\tau_{\mathrm{ph}}\), parallel mode & 0.39\\
Photon survival \(e^{-T/\tau_{\mathrm{ph}}}\) & 0.68\\
Tick duration \(t_{\mathrm{cell}}\) & 84 ps\\
Non-adiabatic phase error at \(T=20/J_x\) & \(\approx0.2\) rad\\
\end{tabular}
\end{table}

\begin{table}[t]
\centering
\caption{Device scale and readout requirements at the operating point of
Table~\ref{tab:budget_operating}.  The device-scale entries follow from the
heralded momentum width and the group velocity; the scan entries from
Sec.~\ref{sec:sampling}; the acquisition entries from the noise target of
Sec.~\ref{sec:digital_protection}.}
\label{tab:budget_readout}
\begin{tabular}{lc}
Quantity & Value \\
\hline
\multicolumn{2}{l}{\textit{Device scale}}\\
Heralded momentum width \(\Delta k_x\) & 0.041 rad\\
Packet extent \(\sim1/\Delta k_x\) & \(\approx25\) cells\\
Ballistic drift per cycle & \(\approx9.5\) cells\\
Chain length \(N\) per arm & \(\gtrsim100\) cells\\
Number of chains, parallel mode & 2\\
Coupler excess loss per arm & \(\approx0.2\) dB\\
\hline
\multicolumn{2}{l}{\textit{Scan requirements}}\\
Peak phase gradient \(\max|\partial_{k_x}\gamma|\) & 4.84\\
Scan settings, minimum / adopted & 20 / 64\\
Filter step spacing \(\delta f\), centre / mean & 0.19 / 0.125 GHz\\
Filter passband, apodised & \(\lesssim0.15\) GHz\\
Largest fringe step \(\Delta_{\max}\) & 0.95 rad\\
\hline
\multicolumn{2}{l}{\textit{Acquisition}}\\
Target fringe phase noise \(\sigma_\varphi\) & 0.4 rad\\
Worst-sample visibility \(V\) & 0.74\\
Coincidences per setting \(N_{cc}\simeq1/(V\sigma_\varphi)^2\) & \(\approx11\)\\
Settings \(\times\) directions \(\times\) \(\theta_{\mathrm{ref}}\) points
& \(64\times2\times4\)\\
Total coincidences required & \(\approx5.6\times10^{3}\)\\
Pair rate, accidentals, total time & see text\\
\end{tabular}
\end{table}

The parallel two-chain implementation of Sec.~\ref{sec:implementation} is the
operative one, since the filtering the momentum scan requires excludes the
time-bin alternative; the lifetime requirement is therefore
\(T<\tau_{\mathrm{ph}}\), which the conservative point meets with a ratio of
\(0.39\) and a photon survival of \(0.68\) per traversal.

Three entries set the device rather than the protocol. The heralded momentum
width fixes the packet extent at roughly \(25\) unit cells, and the group
velocity at the reference point carries that packet about \(9.5\) cells during
one cycle, against a quantised transport of one cell; a chain of at least
\(100\) cells, that is \(200\) rings, follows in each of two nominally identical
arms. Propagation loss over that length is already counted in
\(\tau_{\mathrm{ph}}\), since all rings carry the same loaded \(Q\) and the
amplitude decays with time spent in the chain rather than with cells crossed.
What is not counted there is coupler excess loss, incurred once per coupler
traversed, which at \(0.01\) dB per Mach--Zehnder coupler and roughly ten cells
crossed contributes about \(0.2\) dB per arm.

The acquisition block converts the noise target of
Sec.~\ref{sec:digital_protection} into a count. Shot-noise-limited phase
extraction scales as \(\sigma_\varphi\simeq1/(V\sqrt{N_{cc}})\), so
\(\sigma_\varphi=0.4\) rad at the worst-sample visibility \(V=0.74\) needs about
eleven coincidences per setting; over \(64\) settings, two injection directions
and four \(\theta_{\mathrm{ref}}\) points per setting, this is roughly
\(5.6\times10^{3}\) coincidences in total. The pair generation rate, the
accidentals ratio and hence the total acquisition time depend on the pump power
and the detection chain, and are not fixed by anything else in this budget.

\end{document}